\documentclass[jap,amsmath,amssymb,reprint]{revtex4-1}
\usepackage{graphicx}
\usepackage{dcolumn}
\usepackage{bm}
\newcommand{\gt}{$\gamma^{\prime\prime\prime}$}

\newcommand{\gs}{$\gamma^{\prime}$}

\newcommand{\GMM}{$\alpha^{\prime\prime}\mathrm{-Fe_{16}N_{2}}$}
\newcommand{\eFeyN}{$\varepsilon-\mathrm{Fe_{3-z}N}$}

\begin{document}


\title{Study of magnetic iron nitride thin films deposited by\\ high power impulse magnetron sputtering}

\author{Akhil Tayal$^1$} \author{Mukul
Gupta$^1$}\email{mgupta@csr.res.in/dr.mukul.gupta@gmail.com}
\author{Ajay Gupta$^2$}  \author{V. Ganesan$^1$} \author{Layanta Behera$^1$} \author{Surendra
Singh$^3$} \author{Saibal Basu$^3$}

\affiliation{$^1$UGC-DAE Consortium for Scientific Research,
University Campus, Khandwa Road, Indore-452 001, India\\ $^2$Amity
Center for Spintronic Materials, Amity University, Sector 125,
Noida-201 303, India\\ $^3$Solid State Physics Division, Bhabha
Atomic Research Center, Mumbai-400 085, India}



\begin{abstract}

In this work, we studied phase formation, structural and magnetic
properties of iron-nitride (Fe-N) thin films deposited using high
power impulse magnetron sputtering (HiPIMS) and direct current
magnetron sputtering (dc-MS). The nitrogen partial pressure during
deposition was systematically varied both in HiPIMS and dc-MS.
Resulting Fe-N films were characterized for their microstructure,
magnetic properties and nitrogen concentration. We found that
HiPIMS deposited Fe-N films show a globular nanocrystalline
microstructure and improved soft magnetic properties. In addition,
it was found that the nitrogen reactivity impedes in HiPIMS as
compared to dc-MS. Obtained results can be understood in terms of
distinct plasma properties of HiPIMS.

\end{abstract}



\date{\today}
\maketitle


\section{Introduction}
\label{1}

HiPIMS is a recently developed technique for the deposition of
thin films. Unique plasma conditions associated with it, makes it
a preferred choice over conventional deposition
methods.~\cite{HIPIMS:PRL:Anders,HiPIMS:Gudmundsson,HiPIMS:Anders:JAP:2007,HIPIMS:JMR:Lundin:12}
As compared to dc-MS, the plasma density in HiPIMS is of the order
of 10$^{19}$m$^{-3}$, about 2 orders of magnitude larger than that
in dc-MS.~\cite{HiPIMS:Gudmundsson} In HiPIMS plasma, number of
ionized species exceeds
neutrals.~\cite{HiPIMS:Gudmundsson,HiPIMS:Sarakinos:2010} These
characteristic properties of HiPIMS plasma results in improving
film qualities such as film density, hardness, surface roughness,
better adhesion, dense microstructure etc. Moreover, due to high
metal ionization this technique has a better scope for controlling
film properties deposited via reactive sputtering. As such HiPIMS
has been frequently utilized for the deposition of metal nitrides
such as Al-N,~\cite{HiPIMS:AlN:Guillaumot:2010}
Cr-N,~\cite{HiPIMS:CrN:Ehiasarian:2007,HiPIMS:CrN:Lin:2011,HiPIMS:Hala:CrN:JAP10}
Ti-N,~\cite{HiPIMS:TiN:Lattemann:2010}
Nb-N,~\cite{HIPIMS:Reinhard:CrNNbN:2007} etc. and metal oxide thin
films such as
TiO$_2$,~\cite{HIPIMS:Aiempanakit:2011:TiO2,HiPIMS:TiO2:Konstantinidis:2006}
Al$_2$O$_3$,~\cite{HiPIMS:Al2O3:Wallin:2008,HIPIMS:Wallin:Al2O3:2008}
ZnO,~\cite{HiPIMS:Konstantinidis:ZnO:2007}
ZrO$_2$,~\cite{HIPIMS:ZrO2:TSF14} and
Fe$_2$O$_3$.~\cite{HIPIMS:Fe2O3:CT14,HiPIMS:Fe2O3:TSF13} In these
studies, it was observed that the properties of films deposited
using reactive HiPIMS are superior. Konstantindis $et~al.$ found
that formation of rutile phase in TiO$_2$ thin film is more
favorable as compared to anatase phase when sputtered using
HiPIMS. Moreover, HiPIMS deposited films show higher refractive
index.~\cite{HiPIMS:TiO2:Konstantinidis:2006} Ehiasarian $et~al.$
observed that pretreatment using HiPIMS has improved the adhesion
and mechanical properties of CrN thin
films.~\cite{HiPIMS:CrN:NbN:Ehiasarian:2004} Similarly, Reinhard
$et~al.$ observed improvement in corrosion resistance properties
of HiPIMS treated CrN/NbN superlattice
structure.~\cite{HIPIMS:Reinhard:CrNNbN:2007} Recently, Zhao
$et~al.$ observed that the optical transmittance of Zirconia thin
films deposited using HiPIMS is more as compared to
dc-MS.~\cite{HIPIMS:ZrO2:TSF14}

Looking at the vast capabilities of HiPIMS technique in depositing
various kinds of thin films, it is surprising to note that HiPIMS
processes have not yet been applied for the deposition of magnetic
thin films with recent exception of
Fe$_2$O$_3$~\cite{HIPIMS:Fe2O3:CT14,HiPIMS:Fe2O3:TSF13} and
FeCuNbSiB.~\cite{HIPIMS:FeCuNbSiB:IEEE} Still magnetic nitride
films have not yet been studied with HiPIMS. It is well known that
transition metal magnetic nitrides are an important class of
materials for their usage in various technological
applications.~\cite{Coey.JMMM.1999,PRB:AT:2014} Therefore, it will
be immensely useful to study magnetic nitride films using HiPIMS.

It is well known that Fe-N compounds are interesting both from the
basic and applied point of view. These compounds have a wide range
of usage, such as in tribological coatings, magnetic read-write
heads memory devices,
etc.~\cite{Navio.PRB08,Navio:APL:2009,Liu:APL:2000} In the present
work we deposited a series of Fe-N films using HiPIMS and compared
them with dc-MS. The structure (local and long range), growth and
magnetic properties of the deposited thin films were investigated
using various characterization techniques. We found that HiPIMS
deposited Fe-N films show a globular nanocrystalline
microstructure and improved soft magnetic properties. In addition,
it was found that nitrogen reactivity impedes in HiPIMS as
compared to dc-MS. The obtained results are presented and
discussed in terms of plasma properties of HiPIMS.

\section{Experimental Details}
\label{2}

Fe-N thin films were deposited using HiPIMS and dc-MS techniques
on Si(100) and float glass substrates using a AJA Int. Inc. make
ATC Orion-8 series sputtering system. Pure Fe (purity 99.995\%)
target (diameter 75mm) was sputtered using a mixture of Ar and
N$_2$ gases. Total gas flow was kept constant at 50\,sccm and the
relative partial pressure of nitrogen defined as
R$_{\mathrm{N_2}}$= P$_{\mathrm{N_2}}\times
100\%$/[P$_{\mathrm{N_2}}$+P$_{\mathrm{Ar}}$], (where
P$_{\mathrm{N_2}}$ and P$_{\mathrm{Ar}}$ is nitrogen and argon gas
flow) was kept at 0, 2, 5, 10, 20, 30, 40 and 50. Before
deposition a base pressure of 2$\times 10^{-6}$\,Pa was achieved.
During the deposition pressure was kept constant at 4\,Pa using a
dynamic throttling valve and substrate temperature was kept at
423\,K. For HiPIMS, the deposition parameters used are: peak power
33.3 kW, peak voltage 700\,V, peak current 47\,A, pulse frequency
60\,Hz, pulse duration 150\,$\mu$s and average power 300\,W. For
dc-MS sputtering power was kept at 100\,W. Typical thickness of
deposited films was kept about 80-100\,nm.

Structural characterizations of the samples were carried out with
x-ray diffraction (XRD) using a standard x-ray diffractometer
(Bruker D8 Advance) equipped with Cu\,K-$\alpha$ x-rays source in
$\theta-2\theta$ geometry. Surface morphology was obtained using
atomic force microscopy (AFM) using a Digital Instruments make
Nanoscope E AFM system with a Si$_3$N$_4$ cantilever. Magnetic
properties were studied using a Quantum Design make
superconducting quantum interference device-vibrating sample
magnetometer (S-VSM) and polarized neutron reflectivity (PNR). PNR
measurements were performed at Dhruva Reactor of Bhaba Atomic
Research Center, Mumbai.~\cite{Basu:JNR:PNR:BARC} A magnetic field
of strength 2000\,Oe was applied along the films plane to saturate
samples magnetically. The local structure of nitrogen was studied
using soft x-ray absorption spectroscopy (SXAS) at BL-1 beamline
of Indus-2 synchrotron radiation source at RRCAT,
Indore.~\cite{Phase:SXAS:BL01} SXAS measurements were performed in
an UHV chamber in total electron yield (TEY) mode. For accurate
measurements of nitrogen concentration, $^{14}$N depth profiles
were measured using secondary ion mass spectroscopy (SIMS)
technique (Hiden Analytical SIMS Workstation). For sputtering
O$_2^{+}$ primary ions were used with 5\,keV energy and 400\,nA
beam current. SIMS measurements were performed in an UHV chamber
with a base pressure of the order of 8$\times$10$^{-8}$\,Pa,
during measurements the chamber pressure was
8$\times$10$^{-6}$\,Pa.

\section{Results and Discussion}
\label{3}

\begin{figure} \center
\vspace{-10mm}
\includegraphics [width=80mm,height=65mm] {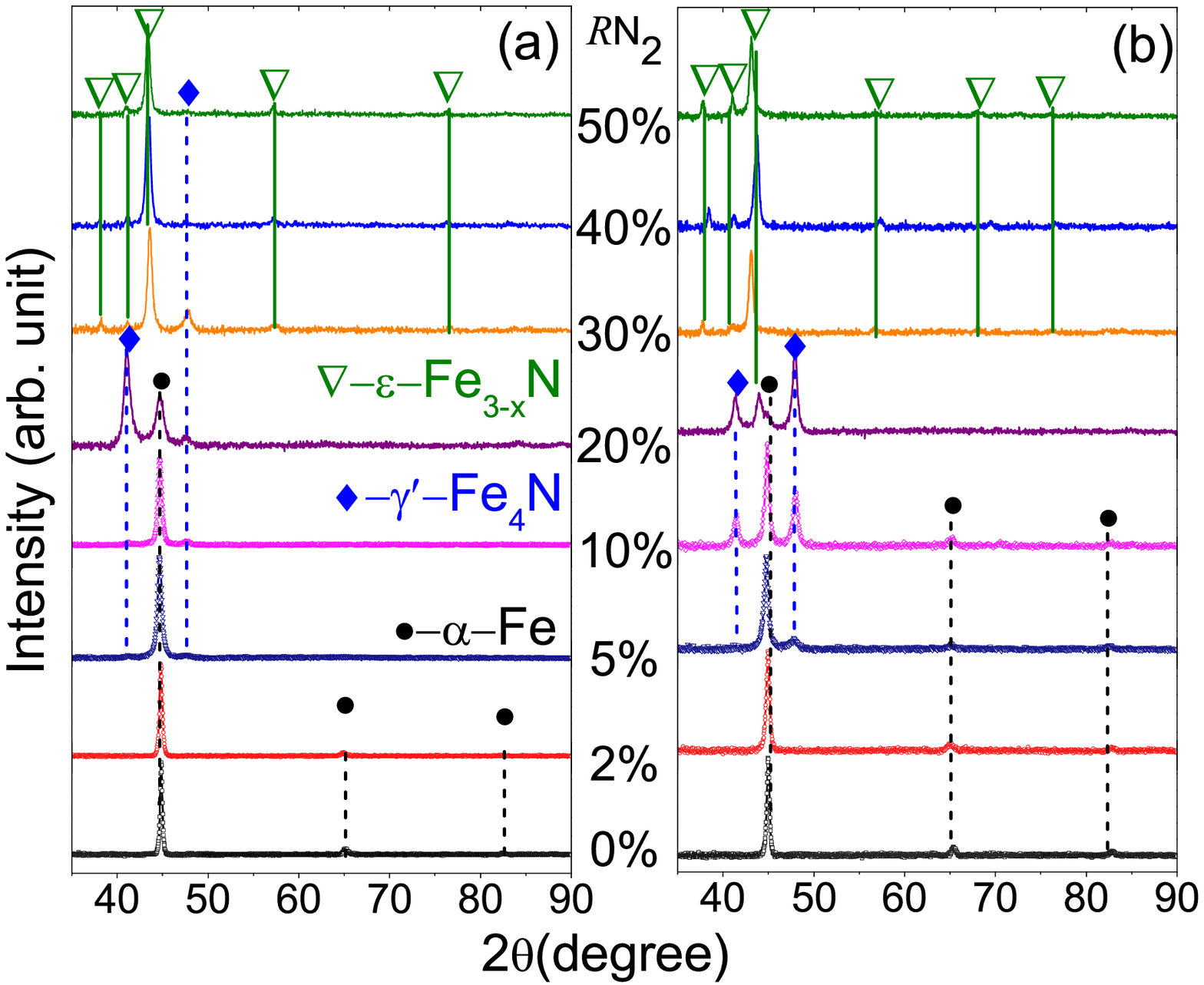}
\caption{\label{fig:XRD}(Color online) X-ray diffraction patterns
of Fe-N thin films prepared at varying R$_{\mathrm{N_2}}$ using
HiPIMS(a) and dc-MS(b) techniques.}
\end{figure}

Phase formation of Fe-N thin films has been extensively studied by
several co-workers using different thin film deposition techniques
such as magnetron sputtering,~\cite{Gupta:PRB05,Easton:TSF05} ion
beam sputtering,~\cite{Borsa.HI.2003,ding:JAP97} e-beam
evaporation,~\cite{Vergnat:TSF96} pulsed laser
deposition,~\cite{Schaaf.PMS.2002,Gupta_JAC01} etc. In general, as
nitrogen partial pressure is increased in a physical vapor
deposition method, different types of Fe-N phases are formed and
they can be broadly classified as: nanocrystalline $\alpha$-Fe-N
$\rightarrow$ amorphous $\alpha$-Fe-N  $\rightarrow$ \GMM
$\rightarrow$ \gs-Fe$_4$N $\rightarrow$ \eFeyN (0$\leq$z$\leq$1)
$\rightarrow$ $\zeta$-Fe$_2$N $\rightarrow$ \gt-FeN $\rightarrow$
amorphous/nanocrystalline \gt-FeN. Since formation of Fe-N phases
with HiPIMS has not yet been studied, we deposited a series of
Fe-N samples using HiPIMS and compared them with its sibling i.e.
dc-MS.

Figure~\ref{fig:XRD} shows XRD patterns of Fe-N thin films
deposited using HiPIMS(a) and dc-MS(b) at different nitrogen
partial pressures. For R$_{\mathrm{N_2}}$=0 and 2\%, the structure
is $bcc~\alpha$-Fe, both with HiPIMS and dc-MS. As
R$_{\mathrm{N_2}}$ increases, $\alpha$-Fe structure is
predominantly preserved up to R$_{\mathrm{N_2}}$ = 10\% in HiPIMS
with faints peaks corresponding to \gs, whereas for dc-MS samples
the amount of \gs~phase appears to be larger. At
R$_{\mathrm{N_2}}$=20\%, intensity of peaks corresponding to
\gs~phase increases in HiPIMS but in dc-MS samples along with
$\alpha$-Fe(N) and \gs~phases, peak corresponding to
$\varepsilon$~phase can also be seen. Further increase in
R$_{\mathrm{N_2}}$ from 30\% to 50\% leads to formation of
$\varepsilon$~phase in both cases, but faint peaks corresponding
to \gs~can only be seen even up to R$_{\mathrm{N_2}}$=50\% in
HiPIMS samples. Observed phase formation with varying
R$_{\mathrm{N_2}}$ is similar in dc-MS to that observed in the
literature, but appears to be somewhat different for HiPIMS
deposited samples.

\begin{figure} \center
\includegraphics [width=58mm,height=40mm] {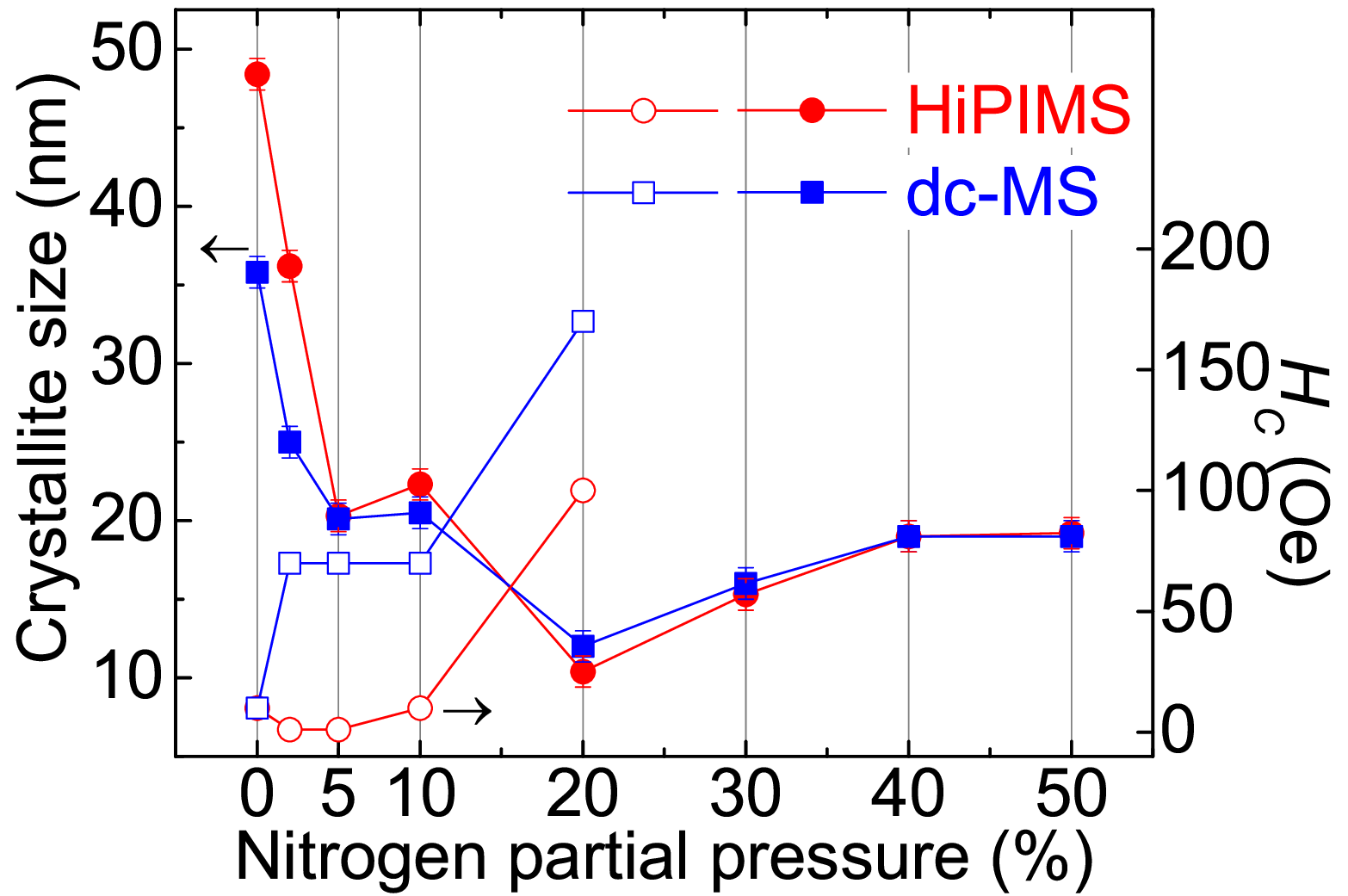}
\caption{\label{fig:GrainSize}(Color online) Variation of
crystallite size and coercivity with increasing nitrogen partial
pressure. Here solid and open symbols represent crystallite size
and coercivity, respectively for HiPIMS and dc-MS samples.}
\end{figure}

\begin{figure*}\center
\includegraphics [width=125mm,height=55mm] {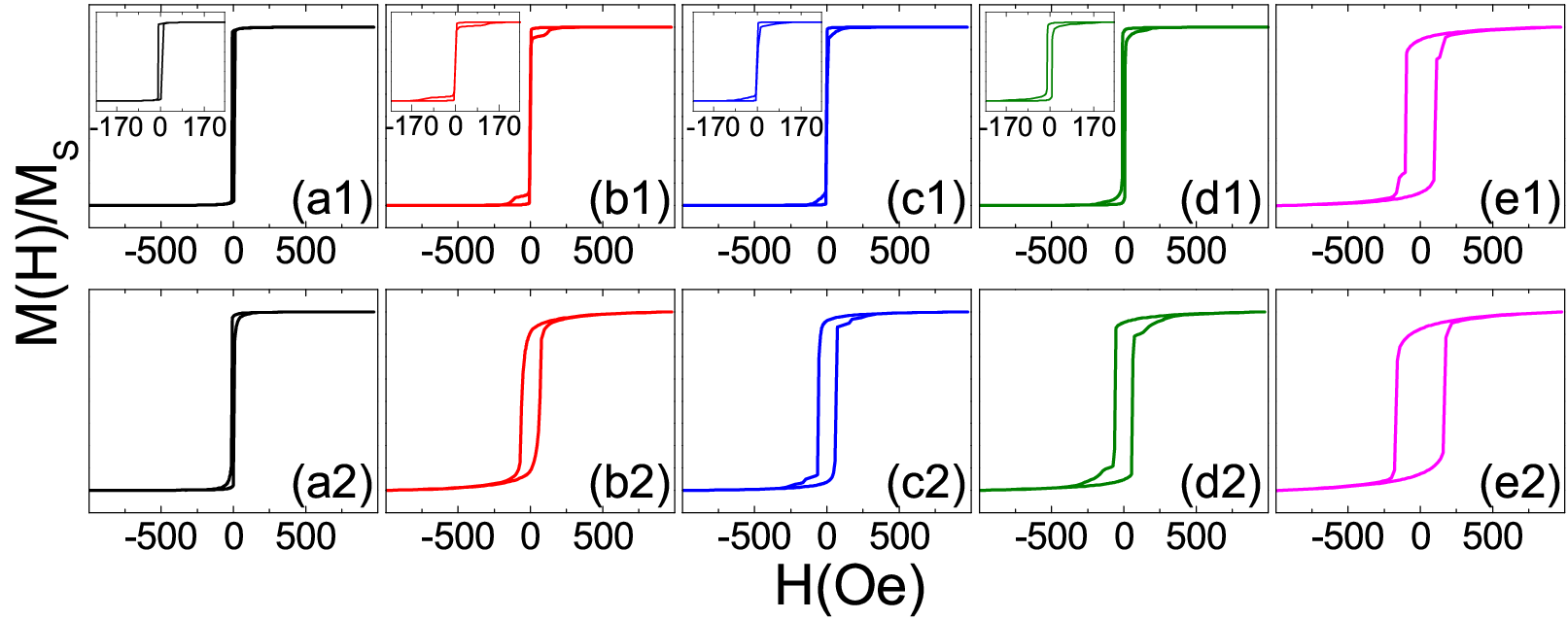}
\caption{\label{fig:mh}(Color online) Normalized M-H loops of Fe-N
thin films deposited using HiPIMS[(a1)-(e1)] and dc-MS[(a2)-(e2)]
at R$_{\mathrm{N_2}}$=0\%[(a1),(a2)], 2\%[(b1),(b2)],
5\%[(c1),(c2)], 10\%[(d1),(d2)], and 20\%[(e1),(e2)]. Inset of
figure[(a1)-(d1)] shows blown up region of M-H loops near the
coercive field.}
\end{figure*}

To see the variation of crystallite size with increasing
R$_{\mathrm{N_2}}$, average crystallite size ($d$) was calculated
using Scherrer formula (for the most intense peak):
$d=0.9\lambda/b\,\cos\theta$~\cite{Cullity_XRD}, and shown in
figure~\ref{fig:GrainSize} [with $\lambda$ wavelength of x-rays,
$b$ an angular width in terms of $2\theta$ and $\theta$ is Bragg
angle]. As such crystallite sizes are similar in both cases except
for R$_{\mathrm{N_2}}$=0 and 2\%, where HiPIMS samples have
significantly larger crystallite sizes as compared with dc-MS.

We did S-VSM and PNR measurements on selected samples to measure
the coercivity ($H_C$) and the saturation magnetization ($M_S$),
respectively. Figure~\ref{fig:mh} shows normalized magnetization
with applied magnetic field (M-H) loops of Fe-N samples deposited
by varying the R$_{\mathrm{N_2}}$ using HiPIMS[(a1)-(e1)] and
dc-MS[(a2)-(e2)] techniques. It can be seen that the M-H loops of
pure Fe films are almost identical  both for HiPIMS and dc-MS,
with $H_C\sim$ 10\,Oe. As nitrogen is introduced during the
deposition, $H_C$ increases suddenly to a value of about 70\,Oe,
for dc-MS samples and remains at this value up to
R$_{\mathrm{N_2}}$=10\%. However, this behavior is strikingly
different for HiPIMS samples; while $H_C$ appears to be negligible
for R$_{\mathrm{N_2}}$ = 2 and 5\% [inset of
figure~\ref{fig:mh}[(a1)-(d1)]] samples, it increases to about
10\,Oe for R$_{\mathrm{N_2}}$=10\%. At R$_{\mathrm{N_2}}$=20\%,
$H_C$ increases in both samples, its value is about 100\,Oe and
170\,Oe for HiPIMS and dc-MS samples, respectively. Typical error
bars in measuring $H_C$ are about $\pm$5\,Oe.

The observed variation in the $H_C$ may stem either due to
particle size effect [explained by random anisotropy model
(RAM)~\cite{Herzer_IEEE89,PRL:Loffler:2000}] or stresses present
in the samples. According to RAM, when particle size is above the
ferromagnetic exchange length($L_{ex}$), the $H_C$ increases with
decreasing the particle size, and for the particle size below
$L_{ex}$, the $H_C$ decreases with decreasing the particle size.
Generally, the value of $L_{ex}\sim$30\,nm, for pure
iron~\cite{Gupta:PRB05,Loffler_PRB98} and expected to vary only
slightly for iron nitride thin films.~\cite{JAP:Bao:RAM:FeN} In
the present case observed variation in the $H_C$ does not
correlate with the change in the crystallite size with increasing
R$_{\mathrm{N_2}}$ [see figure~\ref{fig:GrainSize}]. It indicates
that the observed variance in the $H_C$ can not be explained in
terms of RAM. Therefore, it is expected that the other phenomenon
may be dominant here.

It is known that HiPIMS plasma is highly ionized as compared to
dc-MS even though the energies of adatoms are only slightly higher
in HiPIMS ($\sim$20\,eV). Such conditions lead to denser films
with the relatively smaller amount of defects and vacancies in
HiPIMS as compared to
dc-MS.~\cite{HiPIMS:Gudmundsson,HIPIMS:JMR:Lundin:12,HIPIMS:JPDAP_14:GlobularMs}
Relatively smaller density of defects could result in smaller
stress in the deposited films. This effect lead to improve
soft-magnetic properties of films deposited using HiPIMS. The
sudden increase in $H_C$ for R$_{\mathrm{N_2}}$ =20\%, can be
understood in terms of change in the crystal structure as
discussed in our XRD results. It may be noted that M-H loops of
HiPIMS samples are somewhat asymmetric, which could be due to
growth of \gs~along with $\alpha$-Fe phase, also inferred with XRD
data.

\begin{figure} \center
\includegraphics [width=70mm,height=65mm] {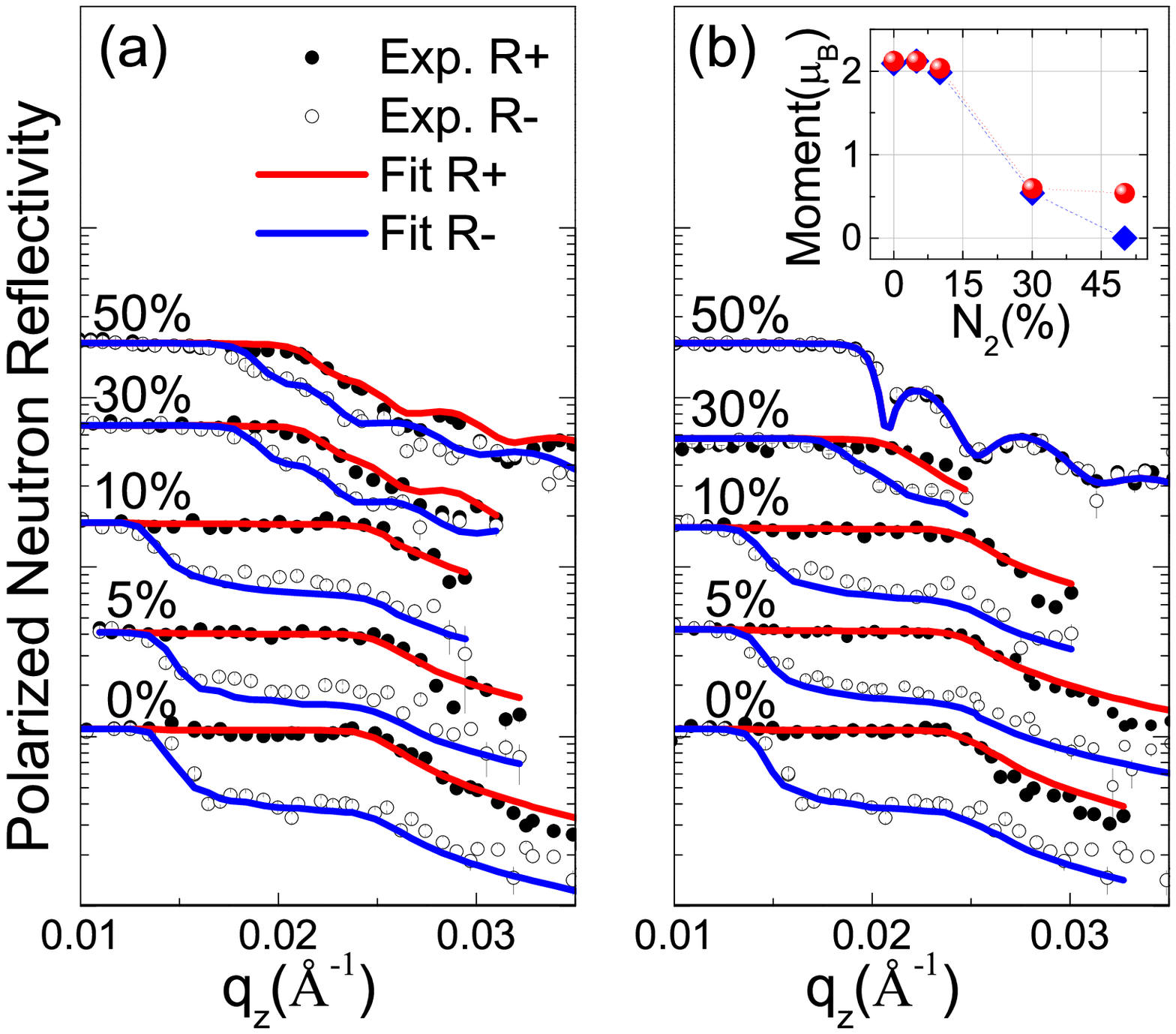}
\caption{\label{fig:pnr}(Color online) PNR patters of Fe-N thin
films prepared using HiPIMS(a) and dc-MS(b) at varying nitrogen
partial pressure. Inset of figure(b) shows variation of average
magnetic moment with increasing nitrogen partial pressure.}
\end{figure}

\label{AFM}

\begin{figure*} \center
\includegraphics[width=140mm,height=60mm]{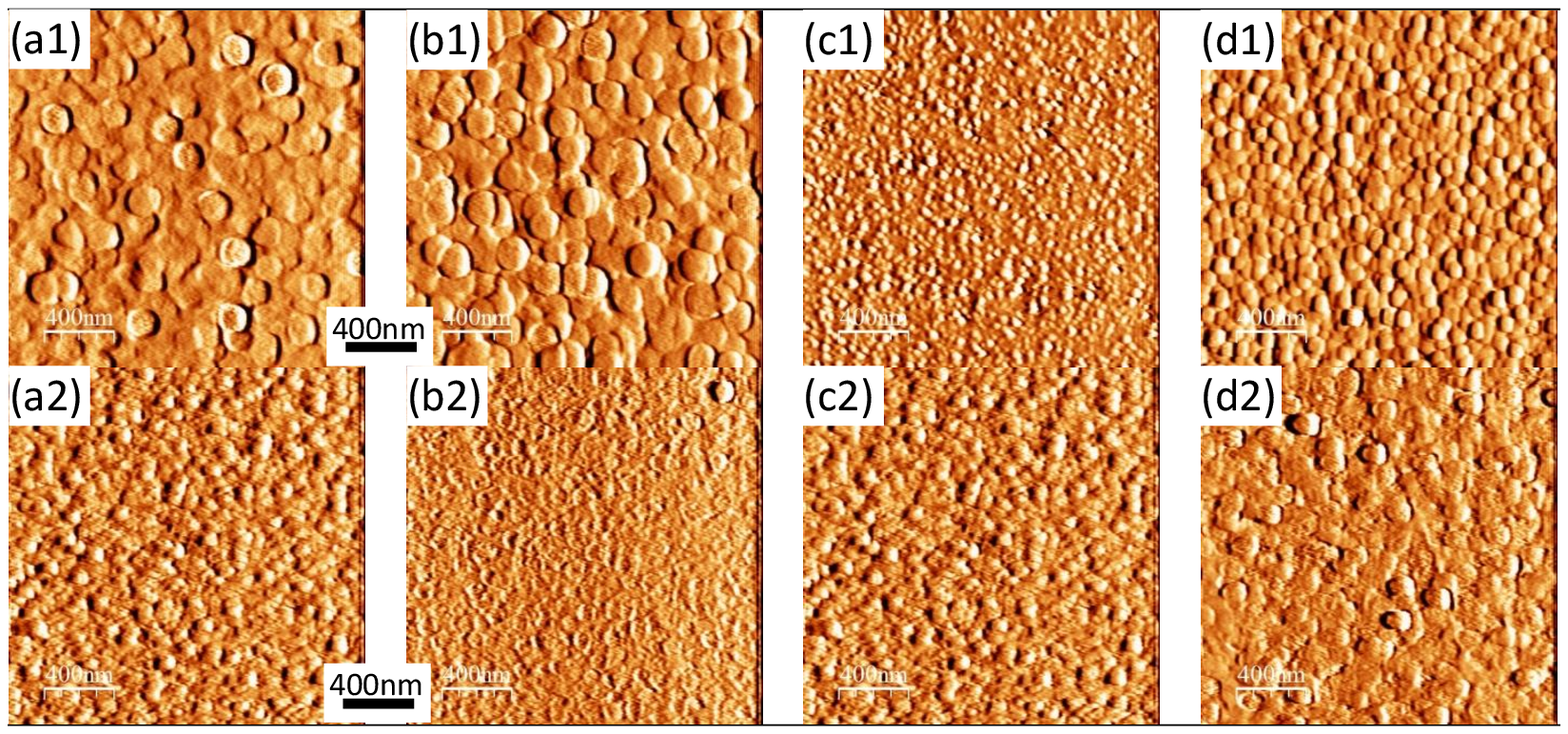}
\caption{\label{fig:AFM}(Color online) AFM images of Fe-N thin
films prepared using HiPIMS[(a1)-(d1)] and dc-MS[(a2)-(d2)] at
$R_{\mathrm{N_2}}$= 0\%[(a1),(a2)], 2\%[(b1),(b2)],
5\%[(c1),(c2)], and 10\%[(d1),(d2)].}
\end{figure*}

To investigate the variation in average magnetic moment with
increasing nitrogen partial pressure, PNR measurements were
performed. It is known that PNR is a very precise tool to measure
accurate average magnetic moment specially for magnetic thin
films, as it is insensitive to thin film
dimensions.~\cite{Blundell_PRB92} Figure~\ref{fig:pnr} shows PNR
patterns of Fe-N thin films deposited using HiPIMS(a) and dc-MS(b)
prepared at varying R$_{\mathrm{N_2}}$. The obtained PNR patterns
were fitted using a computer program based on the Parratt
formalism~\cite{Parratt.PR54} and the average magnetic moment was
calculated. Inset of figure~\ref{fig:pnr}(b) shows the variation
of magnetic moment with increasing R$_{\mathrm{N_2}}$. It can be
seen that up to R$_{\mathrm{N_2}}$=10\%, magnetic moment of the
samples does not show any variation (within experimental accuracy)
both for dc-MS and HiPIMS. However, at R$_{\mathrm{N_2}}$=20\%
there is a sudden decrease in the value of magnetic moment in both
the cases. Interestingly, the sample deposited using dc-MS at
R$_{\mathrm{N_2}}$=50\% becomes non-magnetic, whereas, it remains
magnetic at this nitrogen partial pressure when deposited using
HiPIMS. It is known that the formation of non-magnetic iron
nitrides with increasing nitrogen concentration is only observed
when N concentration
$>$30at.\%.~\cite{Schaaf.PMS.2002,JPC:chen:FexN:1983} As
non-magnetic Fe-N phase is only observed in case of films prepared
using dc-MS, it indicates that nitrogen incorporation has
increased in the samples prepared using dc-MS as compared to
HiPIMS.

The surface morphology of samples was investigated using AFM and
shown in figure~\ref{fig:AFM} for HiPIMS[(a1)-(d1)] and
dc-MS[(a2)-(d2)] at R$_{\mathrm{N_2}}$= 0\%[(a1),(a2)],
2\%[(b1),(b2)], 5\%[(c1),(c2)], and 10\%[(d1),(d2)]. The scan area
for all measurements was kept constant at 2$\mu
m\times\mathrm{2}\mu m$. For HiPIMS samples, the AMF images
indicate that the particle size distribution is more uniform than
that in dc-MS. Recently it was observed that in HiPIMS deposited
films, due to high ionized flux and moderate adatoms energy,
surface mobility increases. This process results in repeated
nucleation process,~\cite{HIPIMS:JMR:Lundin:12} resulting in a
transition from columnar growth (as in dc-MS) to formation of
globular nanocrystalline microstructure as observed in the present
case.~\cite{HIPIMS:JPD:Alami,HIPIMS:JPDAP_14:GlobularMs}

\begin{figure} \center
\includegraphics  [width=80mm,height=35mm] {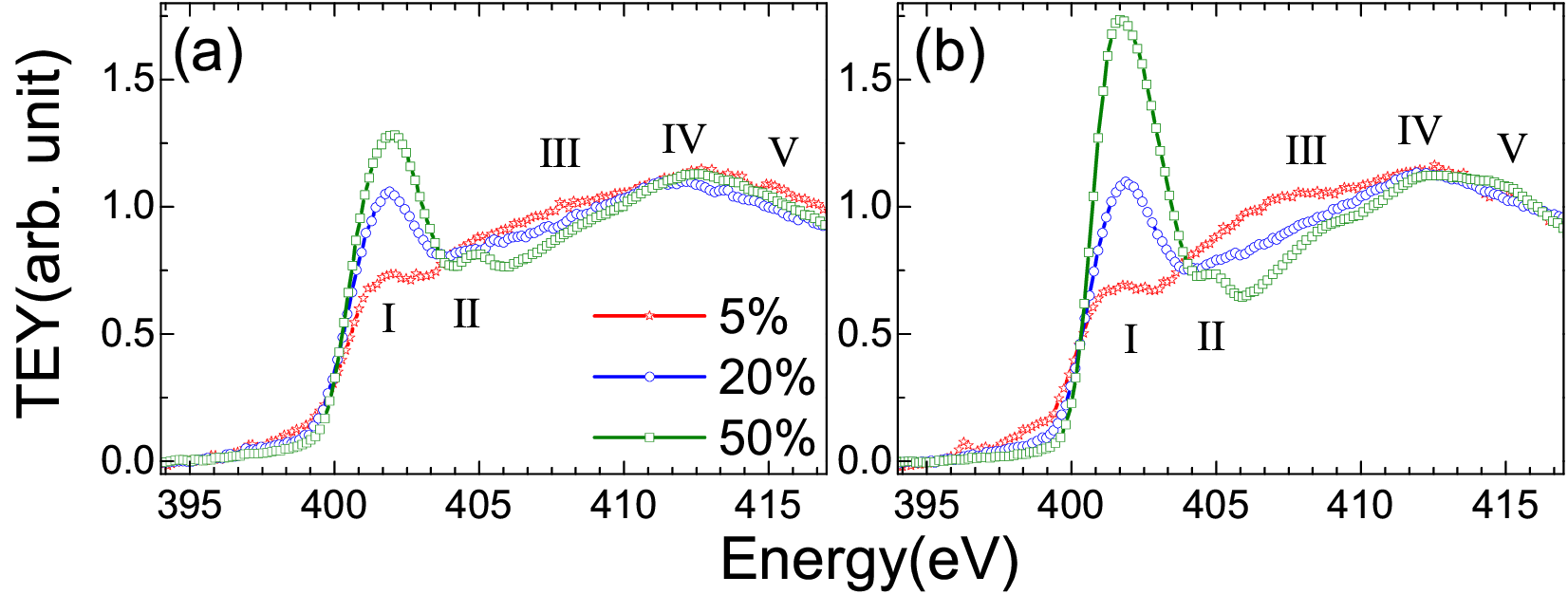}
\caption{\label{fig:XAS}(Color online) Nitrogen K-edge x-ray
absorption spectrum of iron nitride thin films prepared using
HiPIMS(a) and dc-MS(b) at varying R$_{\mathrm{N_2}}$.}
\end{figure}

Soft x-ray absorption spectroscopy is a tool to investigate the
local structure of nitrogen atoms. We did SXAS measurements near
nitrogen K-edge. Obtained spectrum are shown in
figure~\ref{fig:XAS} for selected samples deposited at
R$_{\mathrm{N_2}}$= 5\%, 20\%, and 50\% using HiPIMS(a) and
dc-MS(b). Generally, a SXAS spectrum of metal nitrides consists of
five features as observed in our samples. These features are
assigned as (i) I-transition from N\,$1s$ to the unoccupied
hybridized state of Fe\,$3d-t_{2g}$ and N\,$2p$ (ii) II-transition
from N\,$1s$ to hybridized Fe $3d-e_{g}$ and N $2p$ state (iii)
III, IV,V- transition from N\,$1s$ to hybridized N\,$2p$ and
Fe\,$4sp$
state.~\cite{Pfluger:EELS:89,TSF:CrN:XAS:96,SSC:Mitterbauer:EELS:04,Jouanny2010TSF}
In order to compare the relative nitrogen concentration for HiPIMS
and dc-MS samples, background before pre-edge and after post-edge
was subtracted using a computer program
IFEFFIT-Athena.~\cite{IFEFFIT:Ravel:2005} As can be seen from
figure~\ref{fig:XAS} that the intensity of feature I increases
with increasing R$_{\mathrm{N_2}}$ both for HiPIMS and dc-MS
samples. However, the relative intensity of feature I (for
R$_{\mathrm{N_2}}$=50\%), is significantly enhanced in dc-MS
deposited sample as compared to that in HiPIMS sample. Since the
edge intensity in the XAS pattern is proportional to the nitrogen
concentration,~\cite{EELS:Quantification} the obtained results
clearly indicate that nitrogen concentration is higher for the
samples prepared using dc-MS as compared to HiPIMS.

From the above analysis using various characterization techniques
it appears that the amount of nitrogen in HiPIMS deposited samples
is less than that in dc-MS. In order to quantify N concentration,
we did SIMS measurements. Form SIMS depth profiles, atomic
concentration of a element $A$ in the matrix of element $B$, can
be calculated using a following expression:

\begin{equation}
\label{eq1} C_{A} = RSF_{A} \times \frac{I_A}{I_{B}}
\end{equation}

\begin{figure} \center
\includegraphics  [width=60mm,height=50mm] {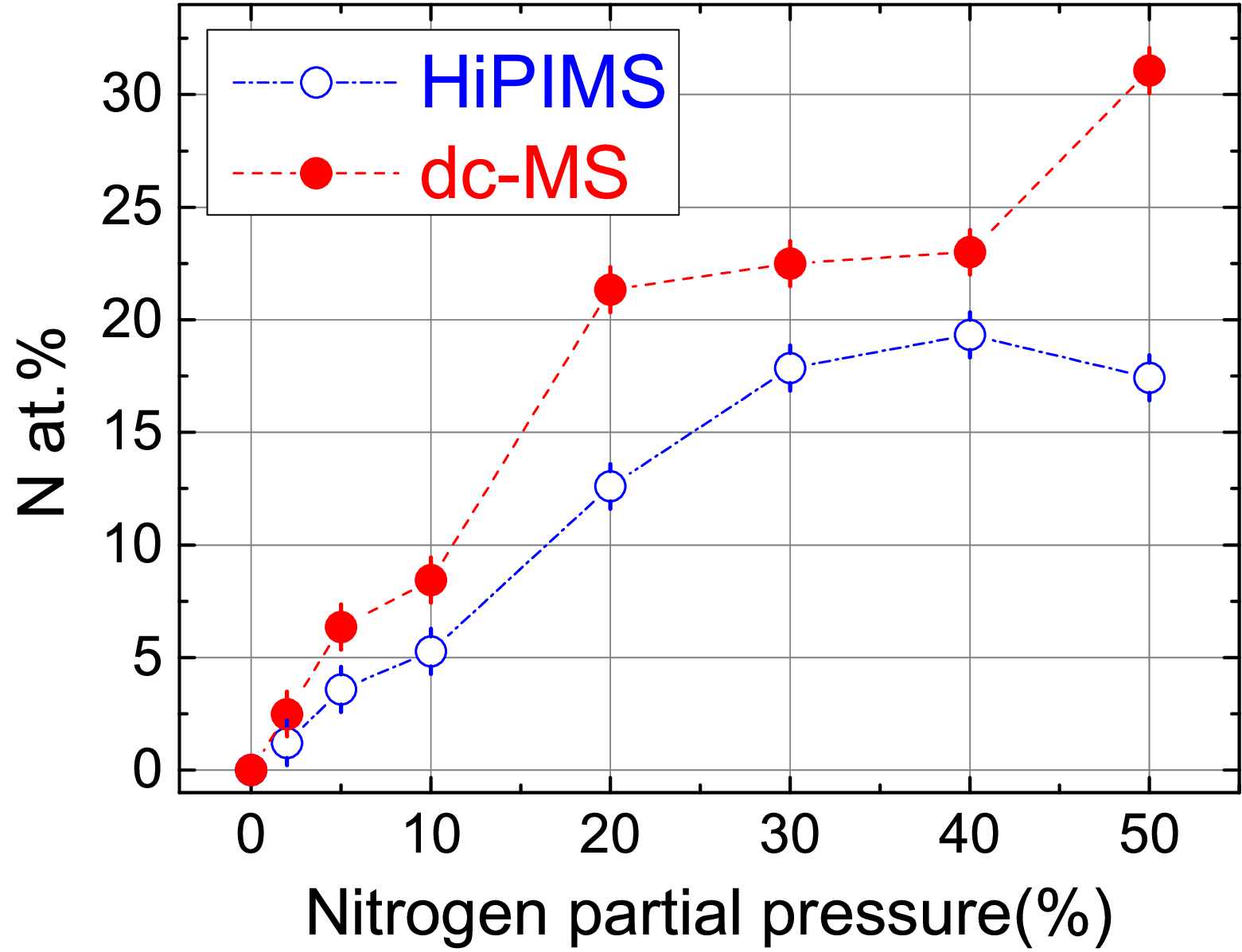}
\caption{\label{fig:SIMS}(Color online) Variation of atomic
concentration of nitrogen with increasing nitrogen partial
pressure deposited using HiPIMS and dc-MS techniques.}
\end{figure}

Where, $C_A$ is atomic concentration, $RSF_A$ is relative
sensitive factor and $I_A$ or $I_B$ is observed intensity in a
SIMS depth profile for element $A$ or $B$. In general, exact
calculation of $RSF$ is a tedious process due to involved matrix
effects, but in the simplest case of two elements $RSF$ can be
calculated by using a reference sample with known
concentration.\cite{PRB:AT:2014,SIMS:Quantification:RCMS10} Using
equation~\ref{eq1} we measured N at.\% both for HiPIMS and dc-MS
samples. Figure~\ref{fig:SIMS} shows the observed variation of
nitrogen concentration with increasing R$_{\mathrm{N_2}}$. It can
be seen that up to R$_{\mathrm{N_2}}$ =20\%, N at.\% increases
almost linearly in both cases. Between R$_{\mathrm{N_2}}$ 20 to
40\%, N at.\% increases marginally but there is a sudden jump at
R$_{\mathrm{N_2}}$=50\% in dc-MS. In contrast, a linear increase
in N at.\% can be seen almost up to R$_{\mathrm{N_2}}$=40\%, and
get saturated for R$_{\mathrm{N_2}}$=50\% in HiPIMS. However,
overall N at.\% is always higher in dc-MS as compared to HiPIMS.

Intuitively, it may appear that the nitrogen concentration should
be more for HiPIMS deposited samples as compared to dc-MS due to
highly ionized flux in HiPIMS. However, as observed in other
nitride systems such as
CrN,~\cite{GasRarefaction:CrN:TiN:JPDAP,gasrarefaction:CrN:Greczynski10}
TiN,~\cite{GasRarefaction:CrN:TiN:JPDAP,Gasrarefaction:TiN:JPDAP}
TiO$_2$,~\cite{GasRarefaction:HIPIMS:Sarakinos07,GasRarefaction:Ludin:12}
etc. the amount of reactive gas get reduced in HiPIMS because of
significantly higher temperature produced in HiPIMS due to high
power impulse (33.3\,kW for 150\,$\mu$s). This leads to expansion
of reactive gas in the vicinity of a target, reducing the volume
of available gas for reaction. This phenomenon is generally known
as
`gas-rarefaction'.~\cite{HiPIMS:Gudmundsson,HIPIMS:JMR:Lundin:12,GasRarefaction:HIPIMS:Sarakinos07,GasRarefaction:Ludin:12}

\section{Conclusion}
\label{Con}

The results obtained from this work provide some distinct
properties of Fe-N films deposited using HiPIMS as compared to
dc-MS deposited films (i) soft magnetic properties Fe-N films
improve (ii) N incorporation in Fe impedes (iii) the
microstructure is globular type rather than columnar. Highly
ionized flux and moderate ion energy in the HiPIMS plasma results
in high adatom mobility leading to the observed effects. A
reduction in N concentration in HiPIMS deposited films as compared
to dc-MS for a similar value of R$_{\mathrm{N_2}}$ can be
understood by `gas-rarefaction' phenomenon caused by substantially
high temperatures in HiPIMS process.

\section*{Acknowledgments} We acknowledge D. M. Phase, D. K. Shukla, R. Sah and S. Karwal for utilization of SXAS
beamline. Help provided in AFM measurements by M. Gangrade, and in
S-VSM measurements by R. J. Choudhary and P. Pandey is gratefully
acknowledged. One of the author (A. T.) wants to acknowledge CSIR,
New Delhi for a research fellowship.

\section*{References}

%


\begin{thebibliography}{10}%
\makeatletter
\providecommand \@ifxundefined [1]{%
 \ifx #1\undefined \expandafter \@firstoftwo
 \else \expandafter \@secondoftwo
\fi
}%
\providecommand \@ifnum [1]{%
 \ifnum #1\expandafter \@firstoftwo
 \else \expandafter \@secondoftwo
\fi
}%
\providecommand \enquote [1]{``#1''}%
\providecommand \bibnamefont  [1]{#1}%
\providecommand \bibfnamefont [1]{#1}%
\providecommand \citenamefont [1]{#1}%
\providecommand\href[0]{\@sanitize\@href}%
\providecommand\@href[1]{\endgroup\@@startlink{#1}\endgroup\@@href}%
\providecommand\@@href[1]{#1\@@endlink}%
\providecommand \@sanitize [0]{\begingroup\catcode`\&12\catcode`\#12\relax}%
\@ifxundefined \pdfoutput {\@firstoftwo}{%
 \@ifnum{\z@=\pdfoutput}{\@firstoftwo}{\@secondoftwo}%
}{%
 \providecommand\@@startlink[1]{\leavevmode}%
 \providecommand\@@endlink[0]{}%
}{%
 \providecommand\@@startlink[1]{%
  \leavevmode
  \pdfstartlink
   attr{/Border[0 0 1 ]/H/I/C[0 1 1]}%
   user{/Subtype/Link/A<</Type/Action/S/URI/URI(#1)>>}%
  \relax
 }%
 \providecommand\@@endlink[0]{\pdfendlink}%
}%
\providecommand \url  [0]{\begingroup\@sanitize \@url }%
\providecommand \@url [1]{\endgroup\@href {#1}{\urlprefix}}%
\providecommand \urlprefix [0]{URL }%
\providecommand \Eprint[0]{\href }%
\@ifxundefined \urlstyle {%
  \providecommand \doi [1]{doi:\discretionary{}{}{}#1}%
}{%
  \providecommand \doi [0]{doi:\discretionary{}{}{}\begingroup
  \urlstyle{rm}\Url }%
}%
\providecommand \doibase [0]{http://dx.doi.org/}%
\providecommand \Doi[1]{\href{\doibase#1}}%
\providecommand \bibAnnote [3]{%
  \BibitemShut{#1}%
  \begin{quotation}\noindent
    \textsc{Key:}\ #2\\\textsc{Annotation:}\ #3%
  \end{quotation}%
}%
\providecommand \bibAnnoteFile [2]{%
  \IfFileExists{#2}{\bibAnnote {#1} {#2} {\input{#2}}}{}%
}%
\providecommand \typeout [0]{\immediate \write \m@ne }%
\providecommand \selectlanguage [0]{\@gobble}%
\providecommand \bibinfo [0]{\@secondoftwo}%
\providecommand \bibfield [0]{\@secondoftwo}%
\providecommand \translation [1]{[#1]}%
\providecommand \BibitemOpen[0]{}%
\providecommand \bibitemStop [0]{}%
\providecommand \bibitemNoStop [0]{.\EOS\space}%
\providecommand \EOS [0]{\spacefactor3000\relax}%
\providecommand \BibitemShut [1]{\csname bibitem#1\endcsname}%
\bibitem{HIPIMS:PRL:Anders}%
  \BibitemOpen
  \bibfield{author}{%
  \bibinfo {author} {\bibfnamefont{J.}~\bibnamefont{Andersson}}\ and\ \bibinfo
  {author} {\bibfnamefont{A.}~\bibnamefont{Anders}},\ }%
  \bibfield{journal}{%
  \Doi{10.1103/PhysRevLett.102.045003}{\bibinfo {journal} {Phys. Rev. Lett.}}\
  }%
  \textbf{\bibinfo {volume} {102}},\ \bibinfo {pages} {045003} (\bibinfo
  {month} {Jan}\ \bibinfo {year} {2009}),\
  \url{http://link.aps.org/doi/10.1103/PhysRevLett.102.045003}%
  \bibAnnoteFile{NoStop}{HIPIMS:PRL:Anders}%
\bibitem{HiPIMS:Gudmundsson}%
  \BibitemOpen
  \bibfield{author}{%
  \bibinfo {author} {\bibfnamefont{J.~T.}\ \bibnamefont{Gudmundsson}}, \bibinfo
  {author} {\bibfnamefont{N.}~\bibnamefont{Brenning}}, \bibinfo {author}
  {\bibfnamefont{D.}~\bibnamefont{Lundin}},\ and\ \bibinfo {author}
  {\bibfnamefont{U.}~\bibnamefont{Helmersson}},\ }%
  \bibfield{journal}{%
  \Doi{http://dx.doi.org/10.1116/1.3691832}{\bibinfo {journal} {Journal of
  Vacuum Science and Technology A}}\ }%
  \textbf{\bibinfo {volume} {30}},\ \bibinfo {eid} {030801} (\bibinfo {year}
  {2012}),\
  \url{http://scitation.aip.org/content/avs/journal/jvsta/30/3/10.1116/1.3691832}%
  \bibAnnoteFile{NoStop}{HiPIMS:Gudmundsson}%
\bibitem{HiPIMS:Anders:JAP:2007}%
  \BibitemOpen
  \bibfield{author}{%
  \bibinfo {author} {\bibfnamefont{A.}~\bibnamefont{Anders}}, \bibinfo {author}
  {\bibfnamefont{J.}~\bibnamefont{Andersson}},\ and\ \bibinfo {author}
  {\bibfnamefont{A.}~\bibnamefont{Ehiasarian}},\ }%
  \bibfield{journal}{%
  \Doi{http://dx.doi.org/10.1063/1.2817812}{\bibinfo {journal} {Journal of
  Applied Physics}}\ }%
  \textbf{\bibinfo {volume} {102}},\ \bibinfo {eid} {113303} (\bibinfo {year}
  {2007}),\
  \url{http://scitation.aip.org/content/aip/journal/jap/102/11/10.1063/1.2817812}%
  \bibAnnoteFile{NoStop}{HiPIMS:Anders:JAP:2007}%
\bibitem{HIPIMS:JMR:Lundin:12}%
  \BibitemOpen
  \bibfield{author}{%
  \bibinfo {author} {\bibfnamefont{D.}~\bibnamefont{Lundin}}\ and\ \bibinfo
  {author} {\bibfnamefont{K.}~\bibnamefont{Sarakinos}},\ }%
  \bibfield{journal}{%
  \bibinfo {journal} {Journal of Materials Research}\ }%
  \textbf{\bibinfo {volume} {27}},\ \bibinfo {pages} {780} (\bibinfo {month}
  {3}\ \bibinfo {year} {2012}),\ ISSN \bibinfo {issn} {2044-5326}%
  \bibAnnoteFile{NoStop}{HIPIMS:JMR:Lundin:12}%
\bibitem{HiPIMS:Sarakinos:2010}%
  \BibitemOpen
  \bibfield{author}{%
  \bibinfo {author} {\bibfnamefont{K.}~\bibnamefont{Sarakinos}}, \bibinfo
  {author} {\bibfnamefont{J.}~\bibnamefont{Alami}},\ and\ \bibinfo {author}
  {\bibfnamefont{S.}~\bibnamefont{Konstantinidis}},\ }%
  \bibfield{journal}{%
  \Doi{http://dx.doi.org/10.1016/j.surfcoat.2009.11.013}{\bibinfo {journal}
  {Surface and Coatings Technology}}\ }%
  \textbf{\bibinfo {volume} {204}},\ \bibinfo {pages} {1661 } (\bibinfo {year}
  {2010}),\ ISSN \bibinfo {issn} {0257-8972},\
  \url{http://www.sciencedirect.com/science/article/pii/S0257897209009426}%
  \bibAnnoteFile{NoStop}{HiPIMS:Sarakinos:2010}%
\bibitem{HiPIMS:AlN:Guillaumot:2010}%
  \BibitemOpen
  \bibfield{author}{%
  \bibinfo {author} {\bibfnamefont{A.}~\bibnamefont{Guillaumot}}, \bibinfo
  {author} {\bibfnamefont{F.}~\bibnamefont{Lapostolle}}, \bibinfo {author}
  {\bibfnamefont{C.}~\bibnamefont{Dublanche-Tixier}}, \bibinfo {author}
  {\bibfnamefont{J.}~\bibnamefont{Oliveira}}, \bibinfo {author}
  {\bibfnamefont{A.}~\bibnamefont{Billard}},\ and\ \bibinfo {author}
  {\bibfnamefont{C.}~\bibnamefont{Langlade}},\ }%
  \bibfield{journal}{%
  \Doi{http://dx.doi.org/10.1016/j.vacuum.2010.04.012}{\bibinfo {journal}
  {Vacuum}}\ }%
  \textbf{\bibinfo {volume} {85}},\ \bibinfo {pages} {120 } (\bibinfo {year}
  {2010}),\ ISSN \bibinfo {issn} {0042-207X},\
  \url{http://www.sciencedirect.com/science/article/pii/S0042207X10001612}%
  \bibAnnoteFile{NoStop}{HiPIMS:AlN:Guillaumot:2010}%
\bibitem{HiPIMS:CrN:Ehiasarian:2007}%
  \BibitemOpen
  \bibfield{author}{%
  \bibinfo {author} {\bibfnamefont{A.~P.}\ \bibnamefont{Ehiasarian}}, \bibinfo
  {author} {\bibfnamefont{J.~G.}\ \bibnamefont{Wen}},\ and\ \bibinfo {author}
  {\bibfnamefont{I.}~\bibnamefont{Petrov}},\ }%
  \bibfield{journal}{%
  \Doi{http://dx.doi.org/10.1063/1.2697052}{\bibinfo {journal} {Journal of
  Applied Physics}}\ }%
  \textbf{\bibinfo {volume} {101}},\ \bibinfo {eid} {054301} (\bibinfo {year}
  {2007}),\
  \url{http://scitation.aip.org/content/aip/journal/jap/101/5/10.1063/1.2697052}%
  \bibAnnoteFile{NoStop}{HiPIMS:CrN:Ehiasarian:2007}%
\bibitem{HiPIMS:CrN:Lin:2011}%
  \BibitemOpen
  \bibfield{author}{%
  \bibinfo {author} {\bibfnamefont{J.}~\bibnamefont{Lin}}, \bibinfo {author}
  {\bibfnamefont{W.~D.}\ \bibnamefont{Sproul}}, \bibinfo {author}
  {\bibfnamefont{J.~J.}\ \bibnamefont{Moore}}, \bibinfo {author}
  {\bibfnamefont{S.}~\bibnamefont{Lee}},\ and\ \bibinfo {author}
  {\bibfnamefont{S.}~\bibnamefont{Myers}},\ }%
  \bibfield{journal}{%
  \Doi{http://dx.doi.org/10.1016/j.surfcoat.2010.11.039}{\bibinfo {journal}
  {Surface and Coatings Technology}}\ }%
  \textbf{\bibinfo {volume} {205}},\ \bibinfo {pages} {3226 } (\bibinfo {year}
  {2011}),\ ISSN \bibinfo {issn} {0257-8972},\
  \url{http://www.sciencedirect.com/science/article/pii/S0257897210012132}%
  \bibAnnoteFile{NoStop}{HiPIMS:CrN:Lin:2011}%
\bibitem{HiPIMS:Hala:CrN:JAP10}%
  \BibitemOpen
  \bibfield{author}{%
  \bibinfo {author} {\bibfnamefont{M.}~\bibnamefont{Hala}}, \bibinfo {author}
  {\bibfnamefont{N.}~\bibnamefont{Viau}}, \bibinfo {author}
  {\bibfnamefont{O.}~\bibnamefont{Zabeida}}, \bibinfo {author}
  {\bibfnamefont{J.~E.}\ \bibnamefont{Klemberg-Sapieha}},\ and\ \bibinfo
  {author} {\bibfnamefont{L.}~\bibnamefont{Martinu}},\ }%
  \bibfield{journal}{%
  \Doi{http://dx.doi.org/10.1063/1.3305319}{\bibinfo {journal} {Journal of
  Applied Physics}}\ }%
  \textbf{\bibinfo {volume} {107}},\ \bibinfo {eid} {043305} (\bibinfo {year}
  {2010}),\
  \url{http://scitation.aip.org/content/aip/journal/jap/107/4/10.1063/1.3305319}%
  \bibAnnoteFile{NoStop}{HiPIMS:Hala:CrN:JAP10}%
\bibitem{HiPIMS:TiN:Lattemann:2010}%
  \BibitemOpen
  \bibfield{author}{%
  \bibinfo {author} {\bibfnamefont{M.}~\bibnamefont{Lattemann}}, \bibinfo
  {author} {\bibfnamefont{U.}~\bibnamefont{Helmersson}},\ and\ \bibinfo
  {author} {\bibfnamefont{J.}~\bibnamefont{Greene}},\ }%
  \bibfield{journal}{%
  \Doi{http://dx.doi.org/10.1016/j.tsf.2010.05.064}{\bibinfo {journal} {Thin
  Solid Films}}\ }%
  \textbf{\bibinfo {volume} {518}},\ \bibinfo {pages} {5978 } (\bibinfo {year}
  {2010}),\ ISSN \bibinfo {issn} {0040-6090},\
  \url{http://www.sciencedirect.com/science/article/pii/S0040609010007340}%
  \bibAnnoteFile{NoStop}{HiPIMS:TiN:Lattemann:2010}%
\bibitem{HIPIMS:Reinhard:CrNNbN:2007}%
  \BibitemOpen
  \bibfield{author}{%
  \bibinfo {author} {\bibfnamefont{C.}~\bibnamefont{Reinhard}}, \bibinfo
  {author} {\bibfnamefont{A.}~\bibnamefont{Ehiasarian}},\ and\ \bibinfo
  {author} {\bibfnamefont{P.}~\bibnamefont{Hovsepian}},\ }%
  \bibfield{journal}{%
  \Doi{http://dx.doi.org/10.1016/j.tsf.2006.11.014}{\bibinfo {journal} {Thin
  Solid Films}}\ }%
  \textbf{\bibinfo {volume} {515}},\ \bibinfo {pages} {3685 } (\bibinfo {year}
  {2007}),\ ISSN \bibinfo {issn} {0040-6090},\
  \url{http://www.sciencedirect.com/science/article/pii/S0040609006013332}%
  \bibAnnoteFile{NoStop}{HIPIMS:Reinhard:CrNNbN:2007}%
\bibitem{HIPIMS:Aiempanakit:2011:TiO2}%
  \BibitemOpen
  \bibfield{author}{%
  \bibinfo {author} {\bibfnamefont{M.}~\bibnamefont{Aiempanakit}}, \bibinfo
  {author} {\bibfnamefont{U.}~\bibnamefont{Helmersson}}, \bibinfo {author}
  {\bibfnamefont{A.}~\bibnamefont{Aijaz}}, \bibinfo {author}
  {\bibfnamefont{P.}~\bibnamefont{Larsson}}, \bibinfo {author}
  {\bibfnamefont{R.}~\bibnamefont{Magnusson}}, \bibinfo {author}
  {\bibfnamefont{J.}~\bibnamefont{Jensen}},\ and\ \bibinfo {author}
  {\bibfnamefont{T.}~\bibnamefont{Kubart}},\ }%
  \bibfield{journal}{%
  \Doi{http://dx.doi.org/10.1016/j.surfcoat.2011.04.071}{\bibinfo {journal}
  {Surface and Coatings Technology}}\ }%
  \textbf{\bibinfo {volume} {205}},\ \bibinfo {pages} {4828 } (\bibinfo {year}
  {2011}),\ ISSN \bibinfo {issn} {0257-8972},\
  \url{http://www.sciencedirect.com/science/article/pii/S0257897211004270}%
  \bibAnnoteFile{NoStop}{HIPIMS:Aiempanakit:2011:TiO2}%
\bibitem{HiPIMS:TiO2:Konstantinidis:2006}%
  \BibitemOpen
  \bibfield{author}{%
  \bibinfo {author} {\bibfnamefont{S.}~\bibnamefont{Konstantinidis}}, \bibinfo
  {author} {\bibfnamefont{J.}~\bibnamefont{Dauchot}},\ and\ \bibinfo {author}
  {\bibfnamefont{M.}~\bibnamefont{Hecq}},\ }%
  \bibfield{journal}{%
  \Doi{http://dx.doi.org/10.1016/j.tsf.2006.07.089}{\bibinfo {journal} {Thin
  Solid Films}}\ }%
  \textbf{\bibinfo {volume} {515}},\ \bibinfo {pages} {1182 } (\bibinfo {year}
  {2006}),\ ISSN \bibinfo {issn} {0040-6090},\ \bibinfo {note} {proceedings of
  the 33rd International Conference on Metallurgical Coatings and Thin Films
  \{ICMCTF\} 2006},\
  \url{http://www.sciencedirect.com/science/article/pii/S0040609006009060}%
  \bibAnnoteFile{NoStop}{HiPIMS:TiO2:Konstantinidis:2006}%
\bibitem{HiPIMS:Al2O3:Wallin:2008}%
  \BibitemOpen
  \bibfield{author}{%
  \bibinfo {author} {\bibfnamefont{E.}~\bibnamefont{Wallin}}\ and\ \bibinfo
  {author} {\bibfnamefont{U.}~\bibnamefont{Helmersson}},\ }%
  \bibfield{journal}{%
  \Doi{http://dx.doi.org/10.1016/j.tsf.2007.08.123}{\bibinfo {journal} {Thin
  Solid Films}}\ }%
  \textbf{\bibinfo {volume} {516}},\ \bibinfo {pages} {6398 } (\bibinfo {year}
  {2008}),\ ISSN \bibinfo {issn} {0040-6090},\
  \url{http://www.sciencedirect.com/science/article/pii/S0040609007015374}%
  \bibAnnoteFile{NoStop}{HiPIMS:Al2O3:Wallin:2008}%
\bibitem{HIPIMS:Wallin:Al2O3:2008}%
  \BibitemOpen
  \bibfield{author}{%
  \bibinfo {author} {\bibfnamefont{E.}~\bibnamefont{Wallin}}, \bibinfo {author}
  {\bibfnamefont{T.~I.}\ \bibnamefont{Selinder}}, \bibinfo {author}
  {\bibfnamefont{M.}~\bibnamefont{Elfwing}},\ and\ \bibinfo {author}
  {\bibfnamefont{U.}~\bibnamefont{Helmersson}},\ }%
  \bibfield{journal}{%
  \bibinfo {journal} {EPL (Europhysics Letters)}\ }%
  \textbf{\bibinfo {volume} {82}},\ \bibinfo {pages} {36002} (\bibinfo {year}
  {2008}),\ \url{http://stacks.iop.org/0295-5075/82/i=3/a=36002}%
  \bibAnnoteFile{NoStop}{HIPIMS:Wallin:Al2O3:2008}%
\bibitem{HiPIMS:Konstantinidis:ZnO:2007}%
  \BibitemOpen
  \bibfield{author}{%
  \bibinfo {author} {\bibfnamefont{S.}~\bibnamefont{Konstantinidis}}, \bibinfo
  {author} {\bibfnamefont{A.}~\bibnamefont{Hemberg}}, \bibinfo {author}
  {\bibfnamefont{J.~P.}\ \bibnamefont{Dauchot}},\ and\ \bibinfo {author}
  {\bibfnamefont{M.}~\bibnamefont{Hecq}},\ }%
  \bibfield{journal}{%
  \bibinfo {journal} {Journal of Vacuum Science and Technology B}\ }%
  \textbf{\bibinfo {volume} {25}} (\bibinfo {year} {2007})%
  \bibAnnoteFile{NoStop}{HiPIMS:Konstantinidis:ZnO:2007}%
\bibitem{HIPIMS:ZrO2:TSF14}%
  \BibitemOpen
  \bibfield{author}{%
  \bibinfo {author} {\bibfnamefont{X.}~\bibnamefont{Zhao}}, \bibinfo {author}
  {\bibfnamefont{J.}~\bibnamefont{Jin}}, \bibinfo {author}
  {\bibfnamefont{J.-C.}\ \bibnamefont{Cheng}}, \bibinfo {author}
  {\bibfnamefont{J.-W.}\ \bibnamefont{Lee}}, \bibinfo {author}
  {\bibfnamefont{K.-H.}\ \bibnamefont{Wu}}, \bibinfo {author}
  {\bibfnamefont{K.-C.}\ \bibnamefont{Lin}}, \bibinfo {author}
  {\bibfnamefont{J.-R.}\ \bibnamefont{Tsai}},\ and\ \bibinfo {author}
  {\bibfnamefont{K.-C.}\ \bibnamefont{Liu}},\ }%
  \bibfield{journal}{%
  \Doi{http://dx.doi.org/10.1016/j.tsf.2014.05.060}{\bibinfo {journal} {Thin
  Solid Films}},\ }%
   (\bibinfo {year} {2014}),\ ISSN \bibinfo {issn} {0040-6090},\
  \url{http://www.sciencedirect.com/science/article/pii/S004060901400635X}%
  \bibAnnoteFile{NoStop}{HIPIMS:ZrO2:TSF14}%
\bibitem{HIPIMS:Fe2O3:CT14}%
  \BibitemOpen
  \bibfield{author}{%
  \bibinfo {author} {\bibfnamefont{S.}~\bibnamefont{Kment}}, \bibinfo {author}
  {\bibfnamefont{Z.}~\bibnamefont{Hubicka}}, \bibinfo {author}
  {\bibfnamefont{J.}~\bibnamefont{Krysa}}, \bibinfo {author}
  {\bibfnamefont{J.}~\bibnamefont{Olejnicek}}, \bibinfo {author}
  {\bibfnamefont{M.}~\bibnamefont{Cada}}, \bibinfo {author}
  {\bibfnamefont{I.}~\bibnamefont{Gregora}}, \bibinfo {author}
  {\bibfnamefont{M.}~\bibnamefont{Zlamal}}, \bibinfo {author}
  {\bibfnamefont{M.}~\bibnamefont{Brunclikova}}, \bibinfo {author}
  {\bibfnamefont{Z.}~\bibnamefont{Remes}}, \bibinfo {author}
  {\bibfnamefont{N.}~\bibnamefont{Liu}}, \bibinfo {author}
  {\bibfnamefont{L.}~\bibnamefont{Wang}}, \bibinfo {author}
  {\bibfnamefont{R.}~\bibnamefont{Kirchgeorg}}, \bibinfo {author}
  {\bibfnamefont{C.}~\bibnamefont{Lee}},\ and\ \bibinfo {author}
  {\bibfnamefont{P.}~\bibnamefont{Schmuki}},\ }%
  \bibfield{journal}{%
  \Doi{http://dx.doi.org/10.1016/j.cattod.2013.11.045}{\bibinfo {journal}
  {Catalysis Today}}\ }%
  \textbf{\bibinfo {volume} {230}},\ \bibinfo {pages} {8 } (\bibinfo {year}
  {2014}),\ ISSN \bibinfo {issn} {0920-5861},\ \bibinfo {note} {selected
  contributions of the 4th International Conference on Semiconductor
  Photochemistry (SP4)},\
  \url{http://www.sciencedirect.com/science/article/pii/S0920586113006500}%
  \bibAnnoteFile{NoStop}{HIPIMS:Fe2O3:CT14}%
\bibitem{HiPIMS:Fe2O3:TSF13}%
  \BibitemOpen
  \bibfield{author}{%
  \bibinfo {author} {\bibfnamefont{Z.}~\bibnamefont{Hubička}}, \bibinfo
  {author} {\bibfnamefont{.}~\bibnamefont{Kment}}, \bibinfo {author}
  {\bibfnamefont{J.}~\bibnamefont{Olejníček}}, \bibinfo {author}
  {\bibfnamefont{M.}~\bibnamefont{Čada}}, \bibinfo {author}
  {\bibfnamefont{T.}~\bibnamefont{Kubart}}, \bibinfo {author}
  {\bibfnamefont{M.}~\bibnamefont{Brunclíková}}, \bibinfo {author}
  {\bibfnamefont{P.}~\bibnamefont{Kšírová}}, \bibinfo {author}
  {\bibfnamefont{P.}~\bibnamefont{Adámek}},\ and\ \bibinfo {author}
  {\bibfnamefont{Z.}~\bibnamefont{Remeš}},\ }%
  \bibfield{journal}{%
  \Doi{http://dx.doi.org/10.1016/j.tsf.2013.09.031}{\bibinfo {journal} {Thin
  Solid Films}}\ }%
  \textbf{\bibinfo {volume} {549}},\ \bibinfo {pages} {184 } (\bibinfo {year}
  {2013}),\ ISSN \bibinfo {issn} {0040-6090},\ \bibinfo {note} {the 40th
  International Conference on Metallurgical Coatings and Thin Films},\
  \url{http://www.sciencedirect.com/science/article/pii/S0040609013014995}%
  \bibAnnoteFile{NoStop}{HiPIMS:Fe2O3:TSF13}%
\bibitem{HiPIMS:CrN:NbN:Ehiasarian:2004}%
  \BibitemOpen
  \bibfield{author}{%
  \bibinfo {author} {\bibfnamefont{A.}~\bibnamefont{Ehiasarian}}, \bibinfo
  {author} {\bibfnamefont{P.}~\bibnamefont{Hovsepian}}, \bibinfo {author}
  {\bibfnamefont{L.}~\bibnamefont{Hultman}},\ and\ \bibinfo {author}
  {\bibfnamefont{U.}~\bibnamefont{Helmersson}},\ }%
  \bibfield{journal}{%
  \Doi{http://dx.doi.org/10.1016/j.tsf.2003.11.113}{\bibinfo {journal} {Thin
  Solid Films}}\ }%
  \textbf{\bibinfo {volume} {457}},\ \bibinfo {pages} {270 } (\bibinfo {year}
  {2004}),\ ISSN \bibinfo {issn} {0040-6090},\
  \url{http://www.sciencedirect.com/science/article/pii/S0040609003017115}%
  \bibAnnoteFile{NoStop}{HiPIMS:CrN:NbN:Ehiasarian:2004}%
\bibitem{HIPIMS:FeCuNbSiB:IEEE}%
  \BibitemOpen
  \bibfield{author}{%
  \bibinfo {author} {\bibfnamefont{I.-L.}\ \bibnamefont{Velicu}}, \bibinfo
  {author} {\bibfnamefont{M.}~\bibnamefont{Neagu}}, \bibinfo {author}
  {\bibfnamefont{H.}~\bibnamefont{Chiriac}}, \bibinfo {author}
  {\bibfnamefont{V.}~\bibnamefont{Tiron}},\ and\ \bibinfo {author}
  {\bibfnamefont{M.}~\bibnamefont{Dobromir}},\ }%
  \bibfield{journal}{%
  \Doi{10.1109/TMAG.2011.2173561}{\bibinfo {journal} {Magnetics, IEEE
  Transactions on}}\ }%
  \textbf{\bibinfo {volume} {48}},\ \bibinfo {pages} {1336} (\bibinfo {month}
  {April}\ \bibinfo {year} {2012}),\ ISSN \bibinfo {issn} {0018-9464}%
  \bibAnnoteFile{NoStop}{HIPIMS:FeCuNbSiB:IEEE}%
\bibitem{Coey.JMMM.1999}%
  \BibitemOpen
  \bibfield{author}{%
  \bibinfo {author} {\bibfnamefont{J.~M.~D.}\ \bibnamefont{Coey}}\ and\
  \bibinfo {author} {\bibfnamefont{P.~A.~I.}\ \bibnamefont{Smith}},\ }%
  \bibfield{journal}{%
  \Doi{DOI: 10.1016/S0304-8853(99)00429-1}{\bibinfo {journal} {J. Magn. Magn.
  Mat.}}\ }%
  \textbf{\bibinfo {volume} {200}},\ \bibinfo {pages} {405 } (\bibinfo {year}
  {1999})%
  \bibAnnoteFile{NoStop}{Coey.JMMM.1999}%
\bibitem{PRB:AT:2014}%
  \BibitemOpen
  \bibfield{author}{%
  \bibinfo {author} {\bibfnamefont{A.}~\bibnamefont{Tayal}}, \bibinfo {author}
  {\bibfnamefont{M.}~\bibnamefont{Gupta}}, \bibinfo {author}
  {\bibfnamefont{N.~P.}\ \bibnamefont{Lalla}}, \bibinfo {author}
  {\bibfnamefont{A.}~\bibnamefont{Gupta}}, \bibinfo {author}
  {\bibfnamefont{M.}~\bibnamefont{Horisberger}}, \bibinfo {author}
  {\bibfnamefont{J.}~\bibnamefont{Stahn}}, \bibinfo {author}
  {\bibfnamefont{K.}~\bibnamefont{Schlage}},\ and\ \bibinfo {author}
  {\bibfnamefont{H.-C.}\ \bibnamefont{Wille}},\ }%
  \bibfield{journal}{%
  \Doi{10.1103/PhysRevB.90.144412}{\bibinfo {journal} {Phys. Rev. B}}\ }%
  \textbf{\bibinfo {volume} {90}},\ \bibinfo {pages} {144412} (\bibinfo {month}
  {Oct}\ \bibinfo {year} {2014}),\
  \url{http://link.aps.org/doi/10.1103/PhysRevB.90.144412}%
  \bibAnnoteFile{NoStop}{PRB:AT:2014}%
\bibitem{Navio.PRB08}%
  \BibitemOpen
  \bibfield{author}{%
  \bibinfo {author} {\bibfnamefont{C.}~\bibnamefont{Nav\'{\i}o}}, \bibinfo
  {author} {\bibfnamefont{J.}~\bibnamefont{Alvarez}}, \bibinfo {author}
  {\bibfnamefont{M.~J.}\ \bibnamefont{Capitan}}, \bibinfo {author}
  {\bibfnamefont{F.}~\bibnamefont{Yndurain}},\ and\ \bibinfo {author}
  {\bibfnamefont{R.}~\bibnamefont{Miranda}},\ }%
  \bibfield{journal}{%
  \Doi{doi:10.1103/PhysRevB.78.155417}{\bibinfo {journal} {Phys. Rev. B}}\ }%
  \textbf{\bibinfo {volume} {78}},\ \bibinfo {pages} {155417} (\bibinfo {year}
  {2008})%
  \bibAnnoteFile{NoStop}{Navio.PRB08}%
\bibitem{Navio:APL:2009}%
  \BibitemOpen
  \bibfield{author}{%
  \bibinfo {author} {\bibfnamefont{C.}~\bibnamefont{Nav\'{\i}o}}, \bibinfo
  {author} {\bibfnamefont{J.}~\bibnamefont{Alvarez}}, \bibinfo {author}
  {\bibfnamefont{M.~J.}\ \bibnamefont{Capitan}}, \bibinfo {author}
  {\bibfnamefont{J.}~\bibnamefont{Camarero}},\ and\ \bibinfo {author}
  {\bibfnamefont{R.}~\bibnamefont{Miranda}},\ }%
  \bibfield{journal}{%
  \Doi{doi:10.1063/1.3159630}{\bibinfo {journal} {Appl. Phys. Lett.}}\ }%
  \textbf{\bibinfo {volume} {94}},\ \bibinfo {pages} {263112} (\bibinfo {year}
  {2009})%
  \bibAnnoteFile{NoStop}{Navio:APL:2009}%
\bibitem{Liu:APL:2000}%
  \BibitemOpen
  \bibfield{author}{%
  \bibinfo {author} {\bibfnamefont{Y.-K.}\ \bibnamefont{Liu}}\ and\ \bibinfo
  {author} {\bibfnamefont{M.~H.}\ \bibnamefont{Kryder}},\ }%
  \bibfield{journal}{%
  \Doi{DOI:10.1063/1.126998}{\bibinfo {journal} {Appl. Phys. Lett.}}\ }%
  \textbf{\bibinfo {volume} {77}},\ \bibinfo {pages} {426} (\bibinfo {year}
  {2000}),\ \url{http://dx.doi.org/doi/10.1063/1.126998}%
  \bibAnnoteFile{NoStop}{Liu:APL:2000}%
\bibitem{Basu:JNR:PNR:BARC}%
  \BibitemOpen
  \bibfield{author}{%
  \bibinfo {author} {\bibfnamefont{S.}~\bibnamefont{Basu}}\ and\ \bibinfo
  {author} {\bibfnamefont{S.}~\bibnamefont{Singh}},\ }%
  \bibfield{journal}{%
  \Doi{10.1080/10238160500475301}{\bibinfo {journal} {Journal of Neutron
  Research}}\ }%
  \textbf{\bibinfo {volume} {14}},\ \bibinfo {pages} {109} (\bibinfo {year}
  {2006}),\
  \Eprint{http://arxiv.org/abs/http://www.tandfonline.com/doi/pdf/10.1080/10238160500475301}{http://www.tandfonline.com/doi/pdf/10.1080/10238160500475301},\
  \url{http://www.tandfonline.com/doi/abs/10.1080/10238160500475301}%
  \bibAnnoteFile{NoStop}{Basu:JNR:PNR:BARC}%
\bibitem{Phase:SXAS:BL01}%
  \BibitemOpen
  \bibfield{author}{%
  \bibinfo {author} {\bibfnamefont{D.~M.}\ \bibnamefont{Phase}}, \bibinfo
  {author} {\bibfnamefont{M.}~\bibnamefont{Gupta}}, \bibinfo {author}
  {\bibfnamefont{S.}~\bibnamefont{Potdar}}, \bibinfo {author}
  {\bibfnamefont{L.}~\bibnamefont{Behera}}, \bibinfo {author}
  {\bibfnamefont{R.}~\bibnamefont{Sah}},\ and\ \bibinfo {author}
  {\bibfnamefont{A.}~\bibnamefont{Gupta}},\ }%
  \bibfield{journal}{%
  \Doi{http://dx.doi.org/10.1063/1.4872719}{\bibinfo {journal} {AIP Conference
  Proceedings}}\ }%
  \textbf{\bibinfo {volume} {1591}},\ \bibinfo {pages} {685} (\bibinfo {year}
  {2014}),\
  \url{http://scitation.aip.org/content/aip/proceeding/aipcp/10.1063/1.4872719}%
  \bibAnnoteFile{NoStop}{Phase:SXAS:BL01}%
\bibitem{Gupta:PRB05}%
  \BibitemOpen
  \bibfield{author}{%
  \bibinfo {author} {\bibfnamefont{R.}~\bibnamefont{Gupta}}\ and\ \bibinfo
  {author} {\bibfnamefont{M.}~\bibnamefont{Gupta}},\ }%
  \bibfield{journal}{%
  \Doi{doi:10.1103/PhysRevB.72.024202}{\bibinfo {journal} {Phys. Rev. B}}\ }%
  \textbf{\bibinfo {volume} {72}},\ \bibinfo {pages} {024202} (\bibinfo {year}
  {2005})%
  \bibAnnoteFile{NoStop}{Gupta:PRB05}%
\bibitem{Easton:TSF05}%
  \BibitemOpen
  \bibfield{author}{%
  \bibinfo {author} {\bibfnamefont{E.~B.}\ \bibnamefont{Easton}}, \bibinfo
  {author} {\bibfnamefont{T.}~\bibnamefont{Buhrmester}},\ and\ \bibinfo
  {author} {\bibfnamefont{J.}~\bibnamefont{Dahn}},\ }%
  \bibfield{journal}{%
  \Doi{10.1016/j.tsf.2005.06.063}{\bibinfo {journal} {Thin Solid Films}}\ }%
  \textbf{\bibinfo {volume} {493}},\ \bibinfo {pages} {60 } (\bibinfo {year}
  {2005}),\ ISSN \bibinfo {issn} {0040-6090},\
  \url{http://www.sciencedirect.com/science/article/pii/S0040609005007030}%
  \bibAnnoteFile{NoStop}{Easton:TSF05}%
\bibitem{Borsa.HI.2003}%
  \BibitemOpen
  \bibfield{author}{%
  \bibinfo {author} {\bibfnamefont{D.~M.}\ \bibnamefont{Borsa}}\ and\ \bibinfo
  {author} {\bibfnamefont{D.~O.}\ \bibnamefont{Boerma}},\ }%
  \bibfield{journal}{%
  \Doi{10.1023/B:HYPE.0000020403.64670.02}{\bibinfo {journal} {Hyp.\,Int.}}\ }%
  \textbf{\bibinfo {volume} {151-152}},\ \bibinfo {pages} {31} (\bibinfo {year}
  {2003})%
  \bibAnnoteFile{NoStop}{Borsa.HI.2003}%
\bibitem{ding:JAP97}%
  \BibitemOpen
  \bibfield{author}{%
  \bibinfo {author} {\bibfnamefont{X.}~\bibnamefont{zhao Ding}}, \bibinfo
  {author} {\bibfnamefont{F.}~\bibnamefont{min Zhang}}, \bibinfo {author}
  {\bibfnamefont{J.}~\bibnamefont{sheng Yan}}, \bibinfo {author}
  {\bibfnamefont{H.}~\bibnamefont{lie Shen}}, \bibinfo {author}
  {\bibfnamefont{X.}~\bibnamefont{Wang}}, \bibinfo {author}
  {\bibfnamefont{X.}~\bibnamefont{huai Liu}},\ and\ \bibinfo {author}
  {\bibfnamefont{D.-F.}\ \bibnamefont{Shen}},\ }%
  \bibfield{journal}{%
  \Doi{10.1063/1.366319}{\bibinfo {journal} {J. Appl. Phys.}}\ }%
  \textbf{\bibinfo {volume} {82}},\ \bibinfo {pages} {5154} (\bibinfo {year}
  {1997})%
  \bibAnnoteFile{NoStop}{ding:JAP97}%
\bibitem{Vergnat:TSF96}%
  \BibitemOpen
  \bibfield{author}{%
  \bibinfo {author} {\bibfnamefont{M.}~\bibnamefont{Vergnat}}, \bibinfo
  {author} {\bibfnamefont{P.}~\bibnamefont{Bauer}}, \bibinfo {author}
  {\bibfnamefont{H.}~\bibnamefont{Chatbi}},\ and\ \bibinfo {author}
  {\bibfnamefont{G.}~\bibnamefont{Marchal}},\ }%
  \bibfield{journal}{%
  \Doi{10.1016/0040-6090(95)07055-9}{\bibinfo {journal} {Thin Solid Films}}\ }%
  \textbf{\bibinfo {volume} {275}},\ \bibinfo {pages} {251 } (\bibinfo {year}
  {1996}),\ ISSN \bibinfo {issn} {0040-6090},\ \bibinfo {note}
  {<ce:title>Papers presented at the European Materials Research Society 1995
  Spring Conference, Symposium E: Structure and Properties of Metallic Thin
  Films and Multilayers</ce:title>},\
  \url{http://www.sciencedirect.com/science/article/pii/0040609095070559}%
  \bibAnnoteFile{NoStop}{Vergnat:TSF96}%
\bibitem{Schaaf.PMS.2002}%
  \BibitemOpen
  \bibfield{author}{%
  \bibinfo {author} {\bibfnamefont{P.}~\bibnamefont{Schaaf}},\ }%
  \bibfield{journal}{%
  \Doi{DOI: 10.1016/S0079-6425(00)00003-7}{\bibinfo {journal} {Prog. Mater.
  Sci.}}\ }%
  \textbf{\bibinfo {volume} {47}},\ \bibinfo {pages} {1 } (\bibinfo {year}
  {2002})%
  \bibAnnoteFile{NoStop}{Schaaf.PMS.2002}%
\bibitem{Gupta_JAC01}%
  \BibitemOpen
  \bibfield{author}{%
  \bibinfo {author} {\bibfnamefont{M.}~\bibnamefont{Gupta}}, \bibinfo {author}
  {\bibfnamefont{A.}~\bibnamefont{Gupta}}, \bibinfo {author}
  {\bibfnamefont{P.}~\bibnamefont{Bhattacharya}}, \bibinfo {author}
  {\bibfnamefont{P.}~\bibnamefont{Misra}},\ and\ \bibinfo {author}
  {\bibfnamefont{L.}~\bibnamefont{Kukreja}},\ }%
  \bibfield{journal}{%
  \Doi{10.1016/S0925-8388(01)01316-0}{\bibinfo {journal} {J. Alloys and
  Compounds}}\ }%
  \textbf{\bibinfo {volume} {326}},\ \bibinfo {pages} {265 } (\bibinfo {year}
  {2001}),\ ISSN \bibinfo {issn} {0925-8388},\ \bibinfo {note}
  {<ce:title>Proceedings of the International Conference on Magnetic Materials
  (ICMM)</ce:title>},\
  \url{http://www.sciencedirect.com/science/article/pii/S0925838801013160}%
  \bibAnnoteFile{NoStop}{Gupta_JAC01}%
\bibitem{Cullity_XRD}%
  \BibitemOpen
  \bibfield{author}{%
  \bibinfo {author} {\bibfnamefont{B.~D.}\ \bibnamefont{Cullity}},\ }%
  \emph{\bibinfo {title} {Elements of X-ray Diffraction}}\ (\bibinfo
  {publisher} {Addison-Wesley,MA},\ \bibinfo {year} {1978})%
  \bibAnnoteFile{NoStop}{Cullity_XRD}%
\bibitem{Herzer_IEEE89}%
  \BibitemOpen
  \bibfield{author}{%
  \bibinfo {author} {\bibfnamefont{G.}~\bibnamefont{Herzer}},\ }%
  \bibfield{journal}{%
  \bibinfo {journal} {I.E.E.E. Trans. Mag.}\ }%
  \textbf{\bibinfo {volume} {25}},\ \bibinfo {pages} {3327} (\bibinfo {year}
  {1989})%
  \bibAnnoteFile{NoStop}{Herzer_IEEE89}%
\bibitem{PRL:Loffler:2000}%
  \BibitemOpen
  \bibfield{author}{%
  \bibinfo {author} {\bibfnamefont{J.~F.}\ \bibnamefont{L\"offler}}, \bibinfo
  {author} {\bibfnamefont{H.-B.}\ \bibnamefont{Braun}},\ and\ \bibinfo {author}
  {\bibfnamefont{W.}~\bibnamefont{Wagner}},\ }%
  \bibfield{journal}{%
  \Doi{10.1103/PhysRevLett.85.1990}{\bibinfo {journal} {Phys. Rev. Lett.}}\ }%
  \textbf{\bibinfo {volume} {85}},\ \bibinfo {pages} {1990} (\bibinfo {month}
  {Aug}\ \bibinfo {year} {2000}),\
  \url{http://link.aps.org/doi/10.1103/PhysRevLett.85.1990}%
  \bibAnnoteFile{NoStop}{PRL:Loffler:2000}%
\bibitem{Loffler_PRB98}%
  \BibitemOpen
  \bibfield{author}{%
  \bibinfo {author} {\bibfnamefont{J.~F.}\ \bibnamefont{Loffler}}, \bibinfo
  {author} {\bibfnamefont{J.~P.}\ \bibnamefont{Meier}}, \bibinfo {author}
  {\bibfnamefont{B.}~\bibnamefont{Doudin}}, \bibinfo {author}
  {\bibfnamefont{J.-P.}\ \bibnamefont{Ansermet}},\ and\ \bibinfo {author}
  {\bibfnamefont{W.}~\bibnamefont{Wagner}},\ }%
  \bibfield{journal}{%
  \bibinfo {journal} {Phys. Rev. B}\ }%
  \textbf{\bibinfo {volume} {57}},\ \bibinfo {pages} {2915} (\bibinfo {year}
  {1998})%
  \bibAnnoteFile{NoStop}{Loffler_PRB98}%
\bibitem{JAP:Bao:RAM:FeN}%
  \BibitemOpen
  \bibfield{author}{%
  \bibinfo {author} {\bibfnamefont{X.}~\bibnamefont{Bao}}, \bibinfo {author}
  {\bibfnamefont{R.~M.}\ \bibnamefont{Metzger}},\ and\ \bibinfo {author}
  {\bibfnamefont{W.~D.}\ \bibnamefont{Doyle}},\ }%
  \bibfield{journal}{%
  \bibinfo {journal} {Journal of Applied Physics}\ }%
  \textbf{\bibinfo {volume} {73}} (\bibinfo {year} {1993})%
  \bibAnnoteFile{NoStop}{JAP:Bao:RAM:FeN}%
\bibitem{HIPIMS:JPDAP_14:GlobularMs}%
  \BibitemOpen
  \bibfield{author}{%
  \bibinfo {author} {\bibfnamefont{E.}~\bibnamefont{Lecoq}}, \bibinfo {author}
  {\bibfnamefont{J.}~\bibnamefont{Guillot}}, \bibinfo {author}
  {\bibfnamefont{D.}~\bibnamefont{Duday}}, \bibinfo {author}
  {\bibfnamefont{J.-B.}\ \bibnamefont{Chemin}},\ and\ \bibinfo {author}
  {\bibfnamefont{P.}~\bibnamefont{Choquet}},\ }%
  \bibfield{journal}{%
  \bibinfo {journal} {Journal of Physics D: Applied Physics}\ }%
  \textbf{\bibinfo {volume} {47}},\ \bibinfo {pages} {195201} (\bibinfo {year}
  {2014}),\ \url{http://stacks.iop.org/0022-3727/47/i=19/a=195201}%
  \bibAnnoteFile{NoStop}{HIPIMS:JPDAP_14:GlobularMs}%
\bibitem{Blundell_PRB92}%
  \BibitemOpen
  \bibfield{author}{%
  \bibinfo {author} {\bibfnamefont{S.~J.}\ \bibnamefont{Blundell}}\ and\
  \bibinfo {author} {\bibfnamefont{J.~A.~C.}\ \bibnamefont{Bland}},\ }%
  \bibfield{journal}{%
  \Doi{10.1103/PhysRevB.46.3391}{\bibinfo {journal} {Phys. Rev. B}}\ }%
  \textbf{\bibinfo {volume} {46}},\ \bibinfo {pages} {3391} (\bibinfo {month}
  {Aug}\ \bibinfo {year} {1992}),\
  \url{http://link.aps.org/doi/10.1103/PhysRevB.46.3391}%
  \bibAnnoteFile{NoStop}{Blundell_PRB92}%
\bibitem{Parratt.PR54}%
  \BibitemOpen
  \bibfield{author}{%
  \bibinfo {author} {\bibfnamefont{L.~G.}\ \bibnamefont{Parratt}},\ }%
  \bibfield{journal}{%
  \bibinfo {journal} {Phys. Rev.}\ }%
  \textbf{\bibinfo {volume} {95}},\ \bibinfo {pages} {359} (\bibinfo {year}
  {1954})%
  \bibAnnoteFile{NoStop}{Parratt.PR54}%
\bibitem{JPC:chen:FexN:1983}%
  \BibitemOpen
  \bibfield{author}{%
  \bibinfo {author} {\bibfnamefont{G.}~\bibnamefont{Chen}}, \bibinfo {author}
  {\bibfnamefont{N.}~\bibnamefont{Jaggi}}, \bibinfo {author}
  {\bibfnamefont{J.}~\bibnamefont{Butt}}, \bibinfo {author}
  {\bibfnamefont{E.}~\bibnamefont{Yeh}},\ and\ \bibinfo {author}
  {\bibfnamefont{L.}~\bibnamefont{Schwartz}},\ }%
  \bibfield{journal}{%
  \bibinfo {journal} {Journal of physical chemistry}\ }%
  \textbf{\bibinfo {volume} {87}},\ \bibinfo {pages} {5326} (\bibinfo {year}
  {1983})%
  \bibAnnoteFile{NoStop}{JPC:chen:FexN:1983}%
\bibitem{HIPIMS:JPD:Alami}%
  \BibitemOpen
  \bibfield{author}{%
  \bibinfo {author} {\bibfnamefont{J.}~\bibnamefont{Alami}}, \bibinfo {author}
  {\bibfnamefont{K.}~\bibnamefont{Sarakinos}}, \bibinfo {author}
  {\bibfnamefont{F.}~\bibnamefont{Uslu}},\ and\ \bibinfo {author}
  {\bibfnamefont{M.}~\bibnamefont{Wuttig}},\ }%
  \bibfield{journal}{%
  \bibinfo {journal} {Journal of Physics D: Applied Physics}\ }%
  \textbf{\bibinfo {volume} {42}},\ \bibinfo {pages} {015304} (\bibinfo {year}
  {2009}),\ \url{http://stacks.iop.org/0022-3727/42/i=1/a=015304}%
  \bibAnnoteFile{NoStop}{HIPIMS:JPD:Alami}%
\bibitem{Pfluger:EELS:89}%
  \BibitemOpen
  \bibfield{author}{%
  \bibinfo {author} {\bibfnamefont{J.}~\bibnamefont{Pfluger}}, \bibinfo
  {author} {\bibfnamefont{J.}~\bibnamefont{Fink}}, \bibinfo {author}
  {\bibfnamefont{G.}~\bibnamefont{Crecelius}}, \bibinfo {author}
  {\bibfnamefont{K.}~\bibnamefont{Bohnen}},\ and\ \bibinfo {author}
  {\bibfnamefont{H.}~\bibnamefont{Winter}},\ }%
  \bibfield{journal}{%
  \Doi{http://dx.doi.org/10.1016/0038-1098(82)90130-2}{\bibinfo {journal}
  {Solid State Communications}}\ }%
  \textbf{\bibinfo {volume} {44}},\ \bibinfo {pages} {489 } (\bibinfo {year}
  {1982}),\ ISSN \bibinfo {issn} {0038-1098},\
  \url{http://www.sciencedirect.com/science/article/pii/0038109882901302}%
  \bibAnnoteFile{NoStop}{Pfluger:EELS:89}%
\bibitem{TSF:CrN:XAS:96}%
  \BibitemOpen
  \bibfield{author}{%
  \bibinfo {author} {\bibfnamefont{F.}~\bibnamefont{Esaka}}, \bibinfo {author}
  {\bibfnamefont{H.}~\bibnamefont{Shimada}}, \bibinfo {author}
  {\bibfnamefont{M.}~\bibnamefont{Imamura}}, \bibinfo {author}
  {\bibfnamefont{N.}~\bibnamefont{Matsubayashi}}, \bibinfo {author}
  {\bibfnamefont{T.}~\bibnamefont{Sato}}, \bibinfo {author}
  {\bibfnamefont{A.}~\bibnamefont{Nishijima}}, \bibinfo {author}
  {\bibfnamefont{A.}~\bibnamefont{Kawana}}, \bibinfo {author}
  {\bibfnamefont{H.}~\bibnamefont{Ichimura}}, \bibinfo {author}
  {\bibfnamefont{T.}~\bibnamefont{Kikuchi}},\ and\ \bibinfo {author}
  {\bibfnamefont{K.}~\bibnamefont{Furuya}},\ }%
  \bibfield{journal}{%
  \Doi{http://dx.doi.org/10.1016/S0040-6090(96)88640-8}{\bibinfo {journal}
  {Thin Solid Films}}\ }%
  \textbf{\bibinfo {volume} {281–282}},\ \bibinfo {pages} {314 } (\bibinfo
  {year} {1996}),\ ISSN \bibinfo {issn} {0040-6090},\
  \url{http://www.sciencedirect.com/science/article/pii/S0040609096886408}%
  \bibAnnoteFile{NoStop}{TSF:CrN:XAS:96}%
\bibitem{SSC:Mitterbauer:EELS:04}%
  \BibitemOpen
  \bibfield{author}{%
  \bibinfo {author} {\bibfnamefont{C.}~\bibnamefont{Mitterbauer}}, \bibinfo
  {author} {\bibfnamefont{C.}~\bibnamefont{HAbert}}, \bibinfo {author}
  {\bibfnamefont{G.}~\bibnamefont{Kothleitner}}, \bibinfo {author}
  {\bibfnamefont{F.}~\bibnamefont{Hofer}}, \bibinfo {author}
  {\bibfnamefont{P.}~\bibnamefont{Schattschneider}},\ and\ \bibinfo {author}
  {\bibfnamefont{H.}~\bibnamefont{Zandbergen}},\ }%
  \bibfield{journal}{%
  \Doi{http://dx.doi.org/10.1016/j.ssc.2004.01.045}{\bibinfo {journal} {Solid
  State Communications}}\ }%
  \textbf{\bibinfo {volume} {130}},\ \bibinfo {pages} {209 } (\bibinfo {year}
  {2004}),\ ISSN \bibinfo {issn} {0038-1098},\
  \url{http://www.sciencedirect.com/science/article/pii/S0038109804000936}%
  \bibAnnoteFile{NoStop}{SSC:Mitterbauer:EELS:04}%
\bibitem{Jouanny2010TSF}%
  \BibitemOpen
  \bibfield{author}{%
  \bibinfo {author} {\bibfnamefont{I.}~\bibnamefont{Jouanny}}, \bibinfo
  {author} {\bibfnamefont{P.}~\bibnamefont{Weisbecker}}, \bibinfo {author}
  {\bibfnamefont{V.}~\bibnamefont{Demange}}, \bibinfo {author}
  {\bibfnamefont{M.}~\bibnamefont{Grafout\'{e}}}, \bibinfo {author}
  {\bibfnamefont{O.}~\bibnamefont{Pe{\~{n}}a}},\ and\ \bibinfo {author}
  {\bibfnamefont{E.}~\bibnamefont{Bauer-Grosse}},\ }%
  \bibfield{journal}{%
  \Doi{10.1016/j.tsf.2009.07.039}{\bibinfo {journal} {Thin Solid Films}}\ }%
  \textbf{\bibinfo {volume} {518}},\ \bibinfo {pages} {1883 } (\bibinfo {year}
  {2010})%
  \bibAnnoteFile{NoStop}{Jouanny2010TSF}%
\bibitem{IFEFFIT:Ravel:2005}%
  \BibitemOpen
  \bibfield{author}{%
  \bibinfo {author} {\bibfnamefont{B.}~\bibnamefont{Ravel}}\ and\ \bibinfo
  {author} {\bibfnamefont{M.}~\bibnamefont{Newville}},\ }%
  \bibfield{journal}{%
  \Doi{10.1107/S0909049505012719}{\bibinfo {journal} {Journal of Synchrotron
  Radiation}}\ }%
  \textbf{\bibinfo {volume} {12}},\ \bibinfo {pages} {537} (\bibinfo {month}
  {Jul}\ \bibinfo {year} {2005}),\
  \url{http://dx.doi.org/10.1107/S0909049505012719}%
  \bibAnnoteFile{NoStop}{IFEFFIT:Ravel:2005}%
\bibitem{EELS:Quantification}%
  \BibitemOpen
  \bibfield{author}{%
  \bibinfo {author} {\bibfnamefont{Y.}~\bibnamefont{Kihn}}, \bibinfo {author}
  {\bibfnamefont{C.}~\bibnamefont{Mirguet}},\ and\ \bibinfo {author}
  {\bibfnamefont{L.}~\bibnamefont{Calmels}},\ }%
  \bibfield{journal}{%
  \Doi{http://dx.doi.org/10.1016/j.elspec.2004.02.170}{\bibinfo {journal}
  {Journal of Electron Spectroscopy and Related Phenomena}}\ }%
  \textbf{\bibinfo {volume} {143}},\ \bibinfo {pages} {117 } (\bibinfo {year}
  {2005}),\ ISSN \bibinfo {issn} {0368-2048},\ \bibinfo {note} {electron Energy
  Loss Spectroscopy in the Electron Microscope Electron Energy Loss
  Spectroscopy in the Electron Microscope},\
  \url{http://www.sciencedirect.com/science/article/pii/S0368204804004098}%
  \bibAnnoteFile{NoStop}{EELS:Quantification}%
\bibitem{SIMS:Quantification:RCMS10}%
  \BibitemOpen
  \bibfield{author}{%
  \bibinfo {author} {\bibfnamefont{M.}~\bibnamefont{Py}}, \bibinfo {author}
  {\bibfnamefont{J.~P.}\ \bibnamefont{Barnes}}, \bibinfo {author}
  {\bibfnamefont{D.}~\bibnamefont{Lafond}},\ and\ \bibinfo {author}
  {\bibfnamefont{J.~M.}\ \bibnamefont{Hartmann}},\ }%
  \bibfield{journal}{%
  \Doi{10.1002/rcm.4904}{\bibinfo {journal} {Rapid Communications in Mass
  Spectrometry}}\ }%
  \textbf{\bibinfo {volume} {25}},\ \bibinfo {pages} {629} (\bibinfo {year}
  {2011}),\ ISSN \bibinfo {issn} {1097-0231},\
  \url{http://dx.doi.org/10.1002/rcm.4904}%
  \bibAnnoteFile{NoStop}{SIMS:Quantification:RCMS10}%
\bibitem{GasRarefaction:CrN:TiN:JPDAP}%
  \BibitemOpen
  \bibfield{author}{%
  \bibinfo {author} {\bibfnamefont{A.}~\bibnamefont{Vetushka}}\ and\ \bibinfo
  {author} {\bibfnamefont{A.~P.}\ \bibnamefont{Ehiasarian}},\ }%
  \bibfield{journal}{%
  \bibinfo {journal} {Journal of Physics D: Applied Physics}\ }%
  \textbf{\bibinfo {volume} {41}},\ \bibinfo {pages} {015204} (\bibinfo {year}
  {2008}),\ \url{http://stacks.iop.org/0022-3727/41/i=1/a=015204}%
  \bibAnnoteFile{NoStop}{GasRarefaction:CrN:TiN:JPDAP}%
\bibitem{gasrarefaction:CrN:Greczynski10}%
  \BibitemOpen
  \bibfield{author}{%
  \bibinfo {author} {\bibfnamefont{G.}~\bibnamefont{Greczynski}}\ and\ \bibinfo
  {author} {\bibfnamefont{L.}~\bibnamefont{Hultman}},\ }%
  \bibfield{journal}{%
  \Doi{http://dx.doi.org/10.1016/j.vacuum.2010.01.055}{\bibinfo {journal}
  {Vacuum}}\ }%
  \textbf{\bibinfo {volume} {84}},\ \bibinfo {pages} {1159 } (\bibinfo {year}
  {2010}),\ ISSN \bibinfo {issn} {0042-207X},\
  \url{http://www.sciencedirect.com/science/article/pii/S0042207X10000795}%
  \bibAnnoteFile{NoStop}{gasrarefaction:CrN:Greczynski10}%
\bibitem{Gasrarefaction:TiN:JPDAP}%
  \BibitemOpen
  \bibfield{author}{%
  \bibinfo {author} {\bibfnamefont{F.}~\bibnamefont{Mitschker}}, \bibinfo
  {author} {\bibfnamefont{M.}~\bibnamefont{Prenzel}}, \bibinfo {author}
  {\bibfnamefont{J.}~\bibnamefont{Benedikt}}, \bibinfo {author}
  {\bibfnamefont{C.}~\bibnamefont{Maszl}},\ and\ \bibinfo {author}
  {\bibfnamefont{A.}~\bibnamefont{von Keudell}},\ }%
  \bibfield{journal}{%
  \bibinfo {journal} {Journal of Physics D: Applied Physics}\ }%
  \textbf{\bibinfo {volume} {46}},\ \bibinfo {pages} {495201} (\bibinfo {year}
  {2013}),\ \url{http://stacks.iop.org/0022-3727/46/i=49/a=495201}%
  \bibAnnoteFile{NoStop}{Gasrarefaction:TiN:JPDAP}%
\bibitem{GasRarefaction:HIPIMS:Sarakinos07}%
  \BibitemOpen
  \bibfield{author}{%
  \bibinfo {author} {\bibfnamefont{K.}~\bibnamefont{Sarakinos}}, \bibinfo
  {author} {\bibfnamefont{J.}~\bibnamefont{Alami}},\ and\ \bibinfo {author}
  {\bibfnamefont{M.}~\bibnamefont{Wuttig}},\ }%
  \bibfield{journal}{%
  \bibinfo {journal} {Journal of Physics D: Applied Physics}\ }%
  \textbf{\bibinfo {volume} {40}},\ \bibinfo {pages} {2108} (\bibinfo {year}
  {2007}),\ \url{http://stacks.iop.org/0022-3727/40/i=7/a=037}%
  \bibAnnoteFile{NoStop}{GasRarefaction:HIPIMS:Sarakinos07}%
\bibitem{GasRarefaction:Ludin:12}%
  \BibitemOpen
  \bibfield{author}{%
  \bibinfo {author} {\bibfnamefont{C.}~\bibnamefont{Huo}}, \bibinfo {author}
  {\bibfnamefont{M.~A.}\ \bibnamefont{Raadu}}, \bibinfo {author}
  {\bibfnamefont{D.}~\bibnamefont{Lundin}}, \bibinfo {author}
  {\bibfnamefont{J.~T.}\ \bibnamefont{Gudmundsson}}, \bibinfo {author}
  {\bibfnamefont{A.}~\bibnamefont{Anders}},\ and\ \bibinfo {author}
  {\bibfnamefont{N.}~\bibnamefont{Brenning}},\ }%
  \bibfield{journal}{%
  \bibinfo {journal} {Plasma Sources Science and Technology}\ }%
  \textbf{\bibinfo {volume} {21}},\ \bibinfo {pages} {045004} (\bibinfo {year}
  {2012}),\ \url{http://stacks.iop.org/0963-0252/21/i=4/a=045004}%
  \bibAnnoteFile{NoStop}{GasRarefaction:Ludin:12}%
\end{thebibliography}

\end{document}